\documentclass[10pt,letterpaper]{article}
\usepackage{amsmath}
\usepackage{amssymb}
\usepackage{graphics}
\usepackage{color}
\usepackage{pictex}
\usepackage{xr-hyper}
\usepackage{ae}
\ifx\pdfoutput\undefined

\usepackage[dvips]{graphicx}
\usepackage
[dvips,
 verbose=false,
 bookmarks=true,
 colorlinks=true,
 linkcolor=webred,
 filecolor=webbrown,
 citecolor=webgreen,
 pagecolor=webblue,
 urlcolor=webblue,
 pdftitle={Fibration structures in toric Calabi-Yau Fourfolds},
 pdfauthor={Falk Rohsiepe},
 pdfsubject={toric geometry, algebraic geometry, string theory,
 string dualities},
 pdfkeywords={toric algebraic geometry string theory dualities},
 bookmarksopen=false,
 pdfpagemode=None,
 pdfview=FitH,
 pdfstartview=FitH,
 extension=pdf]{hyperref}

\else

\usepackage[pdftex]{graphicx}
\usepackage
[pdftex,
 verbose=false,
 bookmarks=true,
 colorlinks=true,
 linkcolor=webred,
 filecolor=webbrown,
 citecolor=webgreen,
 pagecolor=webblue,
 urlcolor=webblue,
 pdftitle={Fibration structures in toric Calabi-Yau Fourfolds},
 pdfauthor={Falk Rohsiepe},
 pdfsubject={toric geometry, algebraic geometry, string theory,
 string dualities},
 pdfkeywords={toric algebraic geometry string theory dualities},
 bookmarksopen=false,
 pdfpagemode=None,
 pdfview=FitH,
 pdfstartview=FitH,
 extension=pdf]{hyperref}
\fi

\definecolor{webred}{rgb}{0,0,0}
\definecolor{webbrown}{rgb}{0,0,0}
\definecolor{webgreen}{rgb}{0,0,0}
\definecolor{webdkgreen}{rgb}{0,0,0}
\definecolor{webblue}{rgb}{0,0,0}

\ifx\forcebwout\undefined
\definecolor{webred}{rgb}{.8,0,0}
\definecolor{webbrown}{rgb}{.6,0,0}
\definecolor{webgreen}{rgb}{0,0.5,0}
\definecolor{webdkgreen}{rgb}{0,0.3,0}
\definecolor{webblue}{rgb}{0,0,0.5}
\fi

\newtheorem{thm}{Theorem}[section]
\newtheorem{cor}[thm]{Corollary}
\newtheorem{lem}[thm]{Lemma}

\newtheorem{defn}[thm]{Definition}
\newtheorem{rem}[thm]{Remark}
\DeclareMathAlphabet\mathscr{U}{eus}{m}{n}
\DeclareMathAlphabet\mathup{OT1}{\rmdefault}{m}{n}

\renewcommand{\[}{\begin{equation}}
\renewcommand{\]}{\end{equation}}

\newcommand{\PP}{\ensuremath{\mathbb{P}}}

\newcommand{\ZZ}{\ensuremath{\mathbb{Z}}}
\newcommand{\QQ}{\ensuremath{\mathbb{Q}}}
\newcommand{\CC}{\ensuremath{\mathbb{C}}}

\newcommand{\RR}{\ensuremath{\mathbb{R}}}
\newcommand{\SSS}{\ensuremath{\mathbb{S}}}

\newcommand{\TTT}{\ensuremath{\mathbb{T}}}

\newcommand{\defas}{\ensuremath{:=}}
\newcommand{\asdef}{\ensuremath{=:}}

\newcommand{\Nabla}{\nabla}

\newcommand{\codim}{\mathrm{codim}}

\newcommand{\threevec}[3]{\left(\begin{smallmatrix}{#1}\\{#2}\\{#3}\end{smallmatrix}\right)}
\newcommand{\twovec}[2]{\binom{#1}{#2}}
\newlength{\colsepsik}

\begin{document}
\rightline{\vbox{\hbox{BONN-TH-2005-01} \hbox{February 2005}}}
\bigskip
\centerline{\LARGE Fibration structures in toric Calabi-Yau
Fourfolds}
\medskip
\centerline{Falk Rohsiepe\footnote{email:
\href{mailto:rohsiepe@th.physik.uni-bonn.de}{rohsiepe@th.physik.uni-bonn.de}}}
\centerline{Physikalisches Institut} \centerline{der Universit\"at
Bonn} \centerline{Nu{\ss}allee 12, D-53115 Bonn}
\centerline{Germany}

\begin{abstract}
In the context of string dualities, fibration structures of
Calabi-Yau manifolds play a prominent role. In particular,
elliptic and K3 fibered Calabi-Yau fourfolds are important for
dualities between string compactifications with four flat
space-time dimensions. A natural framework for studying explicit
examples of such fibrations is given by Calabi-Yau hypersurfaces
in toric varieties, because this class of varieties is
sufficiently large to provide examples with very different
features while still allowing a large degree of explicit control.
In this paper, many examples for elliptic K3 fibered Calabi-Yau
fourfolds are found (not constructed) by searching for reflexive
subpolyhedra of reflexive polyhedra corresponding to hypersurfaces
in weighted projective spaces. Subpolyhedra not always give rise
to fibrations and the obstructions are studied. In addition,
perturbative gauge algebras for dual heterotic string theories are
determined. In order to do so, all elliptically fibered toric K3
surfaces are determined. Then, the corresponding gauge algebras
are calculated without specialization to particular polyhedra for
the elliptic fibers. Finally, the perturbative gauge algebras for
the fourfold fibrations are extracted from the generic fibers and
monodromy.
\end{abstract}
\section{Introduction}
Traditionally, there used to be several different string theories,
all of which had some advantages and some drawbacks. Though by
construction the most unnatural of the string theories, the
heterotic string theories $Het_{E8\times E_8}$ and $Het_{SO(32)}$
appeared as the best candidates for providing a realistic theory
for our world due to the very direct approach of incorporating a
gauge group. This gauge group can then be broken to the gauge
group $SU(3) \times SU(2) \times U(1)$ of the standard model in
the process of compactification to four flat spacetime dimensions.
Geometric compactifications of heterotic string theories are
defined by a Calabi-Yau manifold and a stable vector bundle on
it\footnote{The conditions for these bundles are most easily
solved by embedding the vector bundle connection in the tangent
bundle's connection. For a long time, these were the only known
solutions. They share the common drawback of producing too high a
degree of supersymmetry, namely $N=(2,2)$ worldsheet supersymmetry
leading to $N=2$ space-time supersymmetry. The phenomenologically
desired $N=1$ space-time supersymmetry is obtained by considering
more general bundles leading to $N=(0,2)$ worldsheet
supersymmetry.}. The purely supersymmetric closed string theories
of type IIA and IIB are much more naturally defined\footnote{The
least obvious step in their construction is the GSO projection
\cite{gso1,gso2}, the precise form of which distinguishes between
the two types of theories.} and compactifications do not involve
the complications of specifying a vector bundle in addition to the
Calabi-Yau compactification space. Apart from too high an amount
of supersymmetry, their biggest disadvantage is their lack of
nonabelian gauge groups. In fact, this is only true at first
sight. Nonabelian gauge groups \emph{can} be included by allowing
the compactification spaces to have singularities.

In recent years, the picture has changed by the discovery of a
host of dualities connecting the different types of superstring
theories with each other and also with some new theories. Together
with singular transitions between different compactifications this
opens the dreamer's perspective of dealing with just a single
theory, which by some yet unknown dynamics forces the universe
into its observed state.

The simplest examples of such dualities are T-duality and (not
unrelated) mirror duality. These dualities connect different types
of string theories, namely the type IIA and IIB superstring
theories in their perturbative regime. More general dualities are
not as simple as these and usually even connect one theory in its
perturbative regime with another one in the limit of strong
coupling. Hence, they are only visible when including solitonic
degrees of freedom, namely branes and in particular D-branes.

In addition to dualities between different string theories,
dualities with 11-dimensional supergravity were discovered. This
led to the conjecture of an 11-dimensional quantum theory, from
which all the different string theories and 11-dimensional
supergravity can be obtained as different limits. For this theory,
called M-theory, a perturbative description with membranes as
fundamental objects was proposed\footnote{The letter M later
obtained an additional interpretation, when matrix theories were
suggested as a description for the unifying theory.}.

Yet another player in the duality web is 12-dimensional F-theory.
Leaving open the question, whether such a theory (with
12-dimensional dynamics) really exists, this theory has a sound
foundation as a particular class of type IIB compactifications
with varying dilaton. Such compactifications are classically
impossible and can only be defined by using an $SL(2,\ZZ)$
self-duality of the type IIB string, which is geometrized by
introducing two extra spacetime dimensions. Compactifications of
F-theory are defined by elliptically fibered Calabi-Yau varieties.

A very interesting duality was discovered between F-theory
compactified on an elliptic K3 manifold and the heterotic string
compactified on the two-torus, i.e. the generic fiber. This basic
duality can be lifted to higher dimension of the compactification
spaces by application of fiberwise\footnote{Extrapolating
dualities fiberwise is a somewhat dangerous process. At least, the
interplay between local and global properties of holomorphic
objects forces one to take into account global properties of
fibered spaces, e.g. monodromy. But even if some assumptions (like
the adiabatic argument ``slowly varying'') are unjustified, such
arguments can provide valuable ideas.} dualities. This duality is
particularly appealing, since it offers the perspective of trading
vector bundles for ``just'' an additional complex dimension of the
compactification space.

When studying F-theory compactifications, the special role of the
elliptic fiber (``frozen torus'') may be troublesome. This special
treatment may be circumvented by using another duality, which
connects F-theory on a Calabi-Yau variety $Y$ further compactified
on $\SSS^1$ with M-theory on $Y$. M-theory in turn is believed to
be the strong coupling limit of the type IIA string.

In order to obtain space-time dimension four, the heterotic string
is compactified on a Calabi-Yau threefold and F-theory on an
elliptic Calabi-Yau fourfold. For a fiberwise extension of the
basic duality to a heterotic string theory, the Calabi-Yau
fourfold should be elliptic K3 fibered and the compactification
space of the dual heterotic string theory should be obtained by
replacing the generic K3 fiber with the generic elliptic fiber.

In fact, elliptic K3 fibrations are more than just a tool for
conjecturing dualities. In \cite{needk3fordual} it was shown, that
the compactification space of a type IIA string theory must be a
K3 fibration if it is to be dual to a heterotic string
theory\footnote{One has to be a little bit clearer about what is
meant by ``heterotic string theory''. In this context, one demands
the theory to have a purely geometric phase.}. Note, that when
talking about K3 (elliptic, \ldots) fibrations, I always mean a
fibration with \emph{generic} fiber a K3 surface (elliptic curve,
\ldots).

It is very desirable to have as many explicitly calculable
examples for the objects involved in such dualities as possible.
Such examples are certainly valuable for discovering limitations
of duality hypotheses and might also lead to new unexpected
connections. In addition, the knowledge of sufficiently many
explicit examples might some day help in proving dualities by
application of deformation arguments.

A very powerful tool for constructing interesting yet manageable
examples is given by toric geometry. When toric geometry first was
made widely known to physicist by the work of V.\ Batyrev
\cite{batdualpoly}, it provided the first example of a large class
of threefold varieties which was closed under mirror symmetry,
namely the class of Calabi-Yau hypersurfaces in toric varieties
corresponding to four-dimensional reflexive polyhedra. This class
is a superclass of three-dimensional quasismooth hypersurfaces in
weighted projective spaces. The mirror duality is beautifully
expressed as the combinatorical duality between reflexive
polyhedra. When other dualities than mirror symmetry were
discovered, there was a hope to find equally beautiful
combinatorical representations.

Unfortunately, equally self-contained pictures do not emerge in
more general applications of toric geometry. Toric descriptions
for the vector bundles involved in heterotic string
compactifications did not prove to be very useful. The toric
mirror construction via dual Gorenstein cones
\cite{batbordualcones} describing complete intersection
configurations is a beautiful generalization of the mirror duality
for hypersurfaces, but the class of varieties corresponding to
reflexive Gorenstein cones is much less general than complete
intersection subvarieties of products of weighted projective
spaces\footnote{In particular, the complete intersection
Calabi-Yau varieties in pairs of 0-2 mirror dual heterotic
theories as calculated in \cite{bsw02mirr} almost never are
\emph{both} described by reflexive Gorenstein cones.}.

Turning to four-dimensional spaces used for F- and M-theory
compactifications, the class of transverse hypersurfaces in
weighted projective spaces unfortunately cannot be embedded in the
class of hypersurfaces in toric varieties corresponding to
reflexive polyhedra. But giving up hopes of generality, the class
of four-dimensional Calabi-Yau hypersurfaces in toric varieties
corresponding to reflexive polyhedra provides a
\emph{very}\footnote{The set of inequivalent five-dimensional
reflexive polyhedra, though not completely known, is certainly
larger than the set of five-dimensional weighted projective spaces
allowing for transverse hypersurfaces. Although the class of
families of varieties corresponding to the former is not a
superset of the latter, this is already obvious from trivial
extensions of the 473,800,776 \cite{kreuzscarkeclass4poly}
four-dimensional reflexive polyhedra compared to 1,100,055 weight
sets \cite{wisski}.} large class of examples. These examples allow
the same degree of explicit control as their lower dimensional
counterparts, in particular as far as singularity structures and
counting of moduli are concerned.

As noted above, fibration structures with lower dimensional
Calabi-Yau spaces as generic fibers are very important properties
of compactification spaces. This is even more true in four than in
three dimensions. The question arose, whether such fibrations can
be studied in a framework as simple as the description of
hypersurfaces in toric varieties. Fortunately, the answer is yes
and was found in \cite{searchk3fib}. Since the original treatment
is sufficiently imprecise and original hopes about the generality
of the construction proved to be wrong, I will review the
construction in section \ref{toricfibs}.

In the context of the aforementioned $Het \leftrightarrow F$
duality, elliptic K3 fibered Calabi-Yau hypersurfaces in
five-dimensional toric varieties are of particular interest. Due
to the lack of classification of the corresponding polyhedra, a
class of such varieties was constructed in \cite{klyr} and studied
in e.g. \cite{cy4andf}. Although tailormade for precisely this
application, even in the study of these varieties problems were
encountered concerning the unfulfilled hopes conjectured in
\cite{searchk3fib} --- some of the conjectured fibrations simply
did not exist. This problem was circumvented by passing to other
polyhedra, where the problems did not arise\footnote{There are
examples where inequivalent polyhedra give rise to equivalent
families of hypersurfaces. It was observed, that the Hodge numbers
do not change in the process of passing to the alternative
polyhedron. This suggests, but does not prove, that the
corresponding families of hypersurfaces are identical.}.

In this paper, a large class of examples will be exhibited by
searching for fibration structures rather than constructing them.
Instead of passing to more agreeable polyhedra\footnote{The
existence of suitable replacements is not always clear.}, the
problem concerning the existence of fibrations will be treated by
allowing for more general fibrations. In particular, exceptional
fibers with higher than generic dimension are allowed.

\section{Notational conventions}
I assume that the reader is familiar with basic properties of
toric varieties, reflexive polyhedra and Calabi-Yau hypersurfaces
in toric varieties. Since there are different notational
conventions in use, I will shortly summarize the conventions I use
throughout this paper.

In order not to unnecessarily clobber notation, I will from time
to time refrain from notationally distinguishing between
different, but closely related, objects.

Almost all varieties encountered in this paper will be complex
varieties. Therefore, I will refrain from notationally emphasizing
this and e.g. simply write $\PP^n$ instead of $\CC\PP^n$ for the
complex n-dimensional projective space.

For toric varieties, I start (as usual) with dual lattices $N
\cong \ZZ^n, M = N^\star$. Even if not explicitly stated, I will
always choose fixed isomorphisms $N \cong \ZZ^n \subset \RR^n
\cong N_\RR$ in order to identify linear maps (lattice morphisms)
and real (integer) matrices\footnote{This is not a standard
convention, because for studying abelian quotients of toric
varieties it is more favourable to use different lattices inside
the same real extension.}. This enables me to simply write $N$ for
both the lattice and its real extension.

Fans live in $N$ and will be denoted by the letter $\Sigma$ (no
lattice index due to the fixed isomorphisms). $|\Sigma|$ denotes
the support of $\Sigma$ and $\Sigma^{(d)}$ the subset of $\Sigma$
consisting of cones of dimension $d$. The toric variety
corresponding to a fan $\Sigma$ will be written as $X_\Sigma$ and
$\TTT \subset X_\Sigma$ denotes the open algebraic torus contained
in $X_\Sigma$.

Pairs of dual Polyhedra will be denoted $(\Nabla, \Delta)$,
$\Nabla \subset N$, $\Delta \subset M$. This convention will also
be used for nonreflexive (e.g. non-integral) polyhedra. Inner face
normals will always be normalized to have scalar product $-1$ with
the corresponding hypersurface and can thus (for nonreflexive
polyhedra) fail to be integral.

Given a pair of reflexive Polyhedra $(\Nabla, \Delta)$, $X_\Delta$
will (depending on the context) denote either the set of toric
varieties given by fans over triangulations of $\partial \Nabla$
or a particular element of it. Usually (but not always) the
triangulations will be maximal.

The Calabi-Yau hypersurface defined by the set $Z_p$ of zeroes of
a generic section $p$ of the anticanonical bundle on $X_\Delta$
will be called a \textbf{toric Calabi-Yau hypersurface} (although
it is not a toric variety itself). As usual, the divisors of $Z_p$
given by pullbacks of toric divisors of $X_\Delta$ are called
\textbf{toric divisors}. Divisors not obtained in this way are
called \textbf{nontoric}\footnote{Usually, nontoric divisors are
equivalent to sums of irreducible components of toric divisors.
This statement may fail in the case of K3 hypersurfaces, c.f.
section \ref{ellfibk3}.}.

It is well known, that divisors of $X_\Delta$ corresponding to
rays over points in the interior of a hypersurface of $\Nabla$ do
not intersect the generic Calabi-Yau hypersurface. There are
different ways to deal with such points, in particular when
studying different triangulations of $\partial\Nabla$
corresponding to different cones in the secondary fan. One way is
to omit them introducing singularities in the toric varieties,
which do not not meet the Calabi-Yau hypersurfaces. When studying
K\"ahler moduli spaces and secondary fans, this is the favoured
approach, since one directly obtains the correct dimension for the
space of K\"ahler forms on the hypersurface\footnote{Nevertheless,
different cones in the secondary fan do not necessarily correspond
to topologically different phases of the hypersurface. This
happens, whenever the corresponding triangulations do not differ
on faces of codimension $\ge 2$.}. Another approach is to include
the points in order to obtain maximal triangulations\footnote{This
is technically appealing when e.g.\ studying K3 hypersurfaces.} of
$\partial\Nabla$. Then, the additional degrees of freedom in the
choice of K\"ahler class lie in the kernel of the pullback to the
hypersurface. Yet another way is to completely omit the maximal
dimensional cones from the fan $\Sigma$, which sacrifices
compactness of the toric varieties\footnote{I will use this
approach for the toric prefibrations defined later in this
paper.}. I will write $\partial^1\Nabla$ for the union of faces of
$\Nabla$ with codimension $\ge 2$.

I will freely switch between the classic approach to toric
varieties using fans and lattices and the holomorphic quotient
approach \cite{audin,batyrevholquot,musson,coxholquot}. The
section $p$ defining the hypersurface will be written either in
terms of torus characters or homogenous monomials. For a
description of how to translate between the two representations,
refer to e.g.\ \cite{coxrecent}.

\section{Calabi-Yau fibered toric hypersurfaces}
\label{toricfibs} The main subject of this paper will be (sub-)
families of Calabi-Yau hypersurfaces in toric varieties, the
generic members of which carry a fibration structure with generic
fiber again a Calabi-Yau hypersurface in some lower dimensional
toric variety.

As in the case of the hypersurfaces themselves, one is interested
in a framework, which allows to deduce properties of the generic
fibration structure from the polyhedron describing the embedding
toric variety alone. Such a framework was discovered by the
authors of \cite{searchk3fib}. Put into a nutshell, their
statement was as follows: Such a fibration structure exists,
whenever the dual reflexive polyhedron $\Nabla$ of the embedding
toric variety contains a reflexive subpolyhedron $\Nabla^{(f)}
\subset \Nabla$. The generic fiber then is a Calabi-Yau
hypersurface in $X_{\Delta^{(f)}}$.

As there is no obvious way of making the above statement both
precise and true\footnote{In addition, both the original paper as
well as follow-ups remain unclear about some details of the
fibration structures.}, I will briefly elaborate on this subject.

Let $\Nabla = \Delta^\star$ be a reflexive polyhedron, $X_\Delta$
the toric variety corresponding to some
projective\footnote{Projective triangulations allow strictly
convex functions, which are affine on each simplex. Precisely
under this condition does the toric variety allow a K\"ahler
form.} triangulation of $\partial^1\Nabla$ and $Z_f \subset
X_\Delta$ the Calabi-Yau hypersurface given by the set of zeroes
of the (generic) section $p$ of the anticanonical bundle on
$X_\Delta$. The structures we are looking for are fibrations
$$\begin{array}{ccc}
X_{\Delta^{(f)}} & & \\
\downarrow & & \\
X_\Delta & \stackrel{\pi}{\longrightarrow} & B,
\end{array}$$
where the generic fiber $X_{\Delta^{(f)}}$ is given by a pair
$(\Nabla^{(f)}, \Delta^{(f)})$ of reflexive polyhedra and the base
$B$ of the fibration is again a toric variety. Of course, I demand
$\pi$ to be a morphism of toric varieties. In addition, I require
the intersections $Z_p \cap X_{\Delta^{(f)}}$ to be Calabi-Yau
hypersurfaces over generic points in the base $B$. The toric
fibration alone (i.e. without restricting to $Z_p$) will be called
a \textbf{toric Calabi-Yau prefibration} and will be denoted
$(X_{\Delta^{(f)}}, X_\Delta, B)$.

Let the base $B$ be given by a fan $\Sigma_b$. Corresponding fans,
polyhedra and lattices will be marked with the same letters, e.g.
$\Sigma$ for the fan over some projective triangulation of
$\partial^1\Nabla$, $\Sigma^{(f)}$ for the subfan in $N^{(f)}$,
$\Delta^{(f)} \subset M^{(f)}$, and so on\footnote{Also remember
our convention of writing $N^{(f)}$ for both the lattice and its
real extension.}.

As a toric morphism, $\pi$ is given by a lattice morphism
$\hat\pi: N \twoheadrightarrow N_b$ with the usual conditions
enforced by continuity: $\forall \sigma\in\Sigma\;\exists
\sigma_b\in\Sigma_b:\:\hat\pi(\sigma) \subset \sigma_b$.

It should be noted at this point, that I have fixed a fan $\Sigma$
corresponding to a subfamily of $X_\Delta$. In general, the
corresponding triangulation\footnote{More generally, $\Sigma$ will
not always stem from a triangulation, but rather from a more
general polyhedral partition. In such a case, any simplicial
refinement of $\Sigma$ is also compatible with $\Sigma_b$.} cannot
be replaced by some other triangulation without violating the
conditions on the lattice morphism $\hat\pi$ to define a morphism
of toric varieties. For given $\Nabla$, I will later study whether
appropriate triangulations exists at all.

Note, that there is no need to demand the morphism $\pi$ to be
defined on torus orbits of $X_\Delta$, which do not meet the
hypersurface $Z_p$. I avoid any potential problems in this respect
by using fans $\Sigma$ without maximal dimensional cones, i.e.
fans over triangulations of $\partial^1\Nabla$.

The lattice morphism $\hat\pi$ must be surjective, because
otherwise it could be factored into a surjective morphism followed
by an embedding. The first part would then define a fibration in
its own right (with base given by the same fan but courser
lattice), while the second describes an abelian quotient map from
the intermediate base to the original one. The generic fiber of
the original fibration would then be the disjoint union of the
fibers over multiple points in the intermediate base and hence
could not be of the required type.

Since $B$ is supposed to be the base of the fibration, $\pi$ must
be surjective, or in terms of the fans $\hat\pi(|\Sigma|) =
|\Sigma_b|$. This includes the commonly used case $\Sigma_b =
\hat\pi(\Sigma)$, but does not imply it. As a simple consequence,
$\Sigma_b$ must be complete. This would be obvious if I considered
complete fans $\Sigma$. Here, it follows from
\begin{lem} \label{coverbyeqdimface}
Let $\sigma \subset \QQ^N$ be a strongly convex $N$-dimensional
rational cone, $\sigma' \subset \QQ^n$ a convex rational
$n$-dimensional cone and $\pi: \QQ^N\rightarrow\QQ^n$ a linear map
with $\pi(\sigma) = \sigma'$. Let $\Sigma^{(k)} \defas \{\tau\,
|\, \tau\,\mbox{is a face of}\;\sigma\;\wedge\;\dim\tau = k\}$.
Then $\sigma' = \bigcup_{\tau \in \Sigma^{(n)}} \pi(\tau)$.
\end{lem}
\textbf{Proof:} I will prove $N>n \Rightarrow \sigma' =
\bigcup_{\tau \in \Sigma^{(N-1)}} \pi(\tau)$, which is equivalent
to the assertion by complete induction. I must show, that any
$\pi(x)$, $x\in\sigma$, has a preimage on the boundary of
$\sigma$. Since $N>n$, $\dim \ker \pi \ge 1$ and $\exists\, 0
\not= y \in \ker\pi$. Hence, $\forall\, t\in\QQ: \pi(x+ty) =
\pi(x)$. Since $\sigma$ is strongly convex, the line $x+ty$ must
intersect the boundary of $\sigma$.\hfill$\Box$ \vspace{1ex}

One easily calculates\footnote{using e.g. \CC-valued points} the
preimage of $\TTT_b$ to be the open subvariety of $X_\Sigma$ given
by the fan $\Sigma^{gen} = \{\sigma \cap \ker\hat\pi \mid \sigma
\in \Sigma\}$. Now $\Sigma^{gen} \subset \Sigma$, because strong
convexity of the cones in $\Sigma_b$ implies that $\sigma \cap
\ker\hat\pi$ must be a face of $\sigma$ for any cone $\sigma \in
\Sigma$.

I set $N^{(f)} \defas N \cap \ker\hat\pi$ and $\Sigma^{(f)} =
\{\sigma \cap \ker\hat\pi \mid \sigma \in \Sigma\}$. $N$ can be
split\footnote{c.f.\ appendix \ref{completetobase} for a proof.
Note that the splitting is not unique: We can choose an arbitrary
lift $\tilde{N}_b$ of $N_b$.} as $N = N^{(f)} \oplus \tilde{N}_b$.
Obviously, $\tilde{N}_b \cong N_b$. We have $M =
\left(N^{(f)}\right)^\star \oplus \left(\tilde{N}_b\right)^\star$
with $\left(\tilde{N}_b\right)^\star \cong M_b$ and write $M^{(f)}
\defas \left(N^{(f)}\right)^\star$, $\pi_M: M
\rightarrow M^{(f)}$ for the projection to the second factor and
$L_M: M^{(f)} \rightarrow M$ for the lift induced by our choice of
$\tilde{N}_b$.
 The well known duality between intersections and projections here
manifests itself as $\Delta^{(f)} = \pi_M(\Delta)$.

With this notation,
$$\pi^{-1}(\TTT_b) = X_{\Sigma^{(f)}} \times \TTT_b$$
and $\pi$ operates by projecting to the second factor. The fan for
the generic fiber of the toric prefibration is thus given by
$\Sigma^{(f)}$. If the generic fiber is to be a member of the
family $X_{\Delta^{(f)}}$, we must have $\Nabla^{(f)} = \Nabla
\cap \ker\hat\pi$. Such an embedding $\Nabla^{(f)} \hookrightarrow
\Nabla$ will be called a \textbf{virtual} toric Calabi-Yau
prefibration.

\begin{rem}
It is important, that we consider \emph{generic} fibrations, i.e.\
we allow the fibers to degenerate over subvarieties of strictly
smaller dimension. This is true not only for the Calabi-Yau
fibrations, but also for the prefibrations. The statement found in
\cite{kreuzskarkefib4}, that the toric prefibrations are subject
of an exercise in \cite[p.41]{fulton} is wrong, as in this
reference only locally trivial fibrations are considered and
strong conditions on the fans are derived. These conditions are
not met by (most of) our prefibrations. In fact, interesting
physics (like nonabelian gauge groups) emerges precisely in those
cases, where the fibrations degenerate. As a simple example for
the difference, consider $\PP^1$ fibrations over $\PP^1$: The only
locally trivial fibrations are the Hirzebruch surfaces, while any
blowup thereof yields a generic fibration with bouquets of
$\PP^1$s as fibers over the poles.
\end{rem}

\subsection{Calabi-Yau fibers} \label{cycutout}
Let us assume we have found a toric Calabi-Yau prefibration
$(X_{\Delta^{(f)}}, X_\Delta, B)$ as discussed above. Let $Z_p$ be
the Calabi-Yau hypersurface in $X_\Delta$ given by the set of
zeroes of a generic global section $p$ of the anticanonical
bundle. We will see, that the prefibration makes $Z_p$ a fibration
over the same base. The generic fibers of this fibration are
indeed Calabi-Yau varieties.

We first take a closer look at the intersections of $Z_p$ with the
generic fibers of the prefibration. To this end, we write the
generic section $p$ as
$$p = \sum_{m\in\Delta\cap M} a_m \chi^m,$$
where $a_m \in \CC$ are generic coefficients and $\chi^m$ is the
global section represented by the torus character corresponding to
$m\in M$. This can be split as follows:
\begin{eqnarray}
p & = & \sum_{m\in\Delta\cap M} a_m \chi^m \nonumber \\
 & = & \sum_{m_f \in \Delta_f \cap M^{(f)}}
 \sum_{m\in\Delta\cap M:\;\pi_M(m) = m_f}  a_m \chi^m \nonumber \\
 & = & \sum_{m_f \in \Delta_f \cap M^{(f)}}
 \sum_{m'\in\left(\tilde{N}_b\right)^\star: L_M(m_f)+m'\in\Delta}
 a_m \chi^m \nonumber \\
 & = & \sum_{m_f \in \Delta_f \cap M^{(f)}} \left(
 \sum_{m'\in\left(\tilde{N}_b\right)^\star: L_M(m_f)+m'\in\Delta}
 a_m \chi^{m'} \right) \chi^{L_M(m_f)}. \label{fibercy1}
\end{eqnarray}
Let $\{n_i \in N^{(f)}\}$ denote the primitive generators of the
rays in $\left(\Sigma^{(f)}\right)^{(1)}$. Introducing homogenous
coordinates $X_i, i\in I$ on the fiber space and replacing torus
characters by the corresponding homogenous monomials, the
restriction of $p$ to $X_{\Sigma^{(f)}} \times \TTT_b$ can be
rewritten as \[p \sim \sum_{m_f \in \Delta_f \cap M^{(f)}} \left(
 \sum_{m'\in\left(N_b\right)^\star: L_M(m_f) + m'\in\Delta}
 a_m \chi^{m'} \right) \prod_{i\in I} X_i^{\langle n_i, m_f
 \rangle + 1}.\label{fibercy2}\]
Over any point in the base $\TTT_b$ (\ref{fibercy1}) or
(\ref{fibercy2}) is a global section of the anticanonical bundle
on the fiber space, whose coefficients are given by Laurent
polynomials
 $$A_{m_f} \defas \sum_{m'\in\left(\tilde{N}_b\right)^\star: L_M(m_f)+m'\in\Delta}
 a_m \chi^{m'}$$
in the base torus coordinates. At least over generic points in
$\TTT_b$ one has $p \not\equiv 0$ and thus the intersection of the
generic fiber with the Calabi-Yau hypersurface is indeed a
Calabi-Yau variety.

We still need to check, whether $X_{\Sigma_b}$ is the base not
only of our toric prefibration, but also of the Calabi-Yau
fibration, i.e. whether $\pi$ remains surjective when restricted
to the Calabi-Yau hypersurface $Z_p$. The calculations done so far
show, that $\pi(Z_p)$ is a Zariski open subset of $X_b$. As $Z_p$
is compact and $\pi$ is continuous, $\pi(Z_p)$ is compact as
well\footnote{Here we use the analytic topology.}. Hence,
$\pi(Z_p) = B$.

\subsubsection{How generic are the fibers?}
Most of the theorems yielding properties of Calabi-Yau
hypersurfaces from those of the embedding toric variety demand the
defining section to be generic. Hence, it must be ensured that the
(generically chosen) section $p$ remains sufficiently generic when
restricted to the prefibration fibers.

Obviously, the only case for which a coefficient $A_{m_f}$ will
not yield generic values over generic points in the base is
$A_{m_f} \equiv 0$. Unfortunately, this can happen. For a simple
example consider the three-dimensional reflexive Polyhedron
$\Delta_3$ with vertices $\threevec{1}{0}{0}, \threevec{0}{1}{0},
\threevec{-1}{-2}{-1}, \threevec{0}{1}{1}$ and inner face normals
$\threevec{-1}{-1}{4}$, $\threevec{-1}{-1}{0}$,
$\threevec{-1}{3}{-4}$ and $\threevec{3}{-1}{0}$. Projecting along
the direction of the third coordinate, we obtain the reflexive
triangle with vertices $\twovec{1}{0}, \twovec{0}{1}$ and
$\twovec{-1}{-2}$. Apart from the vertices and the origin, this
contains one additional integer point, namely $\twovec{0}{-1}$,
which has no integer preimage in $\Delta_3$.

The good news is, that this does not endanger
$\Sigma^{(f)}$-regularity of the generic Calabi-Yau fiber, since
the Bertini-type argument already works when only using the
monomials for the vertices and the origin contained in
$\Delta^{(f)}$. These in turn always have integer preimages due to
simple convexity arguments.

Unfortunately, for deducing the Picard lattice of a generic K3
fiber, $\Sigma$-regularity is not sufficient \cite{latticek3}. If
the projection map $$\pi_M:\,\Delta\cap M \longrightarrow
\Delta^{(f)} \cap M_f$$ is not surjective, toric methods might
only reveal a sublattice of the true generic Picard
lattice\footnote{This loss of knowledge is certainly not
satisfactory, but it is not easily resolved. Lower bounds on the
enhanced Picard lattices could be obtained e.g. by studying
enlarged automorphism groups of the hypersurfaces.}.

\subsection{Obstructions} \label{obsdisc}
As remarked above, the existence of a reflexive subpolyhedron in
general is not sufficient for finding toric Calabi-Yau
prefibrations (and corresponding toric Calabi-Yau fibrations). The
additional requirement is the existence of fans $\Sigma,
\Sigma_b$, for which the lattice morphism dictated by the
subpolyhedron defines a toric morphism.

Such fans always exist, if the codimension of $\Nabla^{(f)}
\subset \Nabla$ is one. In this case, $\Nabla^{(f)}$ separates
$\partial\Nabla$ into two halves, which can always be triangulated
separately. Any such triangulation defines a fan compatible with
the fan of $\PP^1$, which is the only possibility in this case.

Let us now turn to the case of higher codimension. If we consider
some given triangulation of $\Nabla$ resp.\ the fan $\Sigma$ given
by it, it is easy to check whether there exists a compatible fan
$\Sigma_b$. We start with the set of cones given by
$\tilde\Sigma_b \defas \hat\pi(\Sigma)$. In most cases, this will
not be a fan, but since we want $\hat\pi$ to define a toric
morphism, the fan $\Sigma_b$ must satisfy
$$\forall \tilde\sigma \in\tilde\Sigma_b \; \exists \sigma\in\Sigma_b:
\:\tilde\sigma \subset \sigma.$$

If $\tilde\Sigma_b$ contains a cone, which is not \emph{strongly}
convex, a compatible fan obviously cannot exist. Otherwise,
$\tilde\Sigma_b$ can only fail to be a fan in its own right due to
the existence of cones, whose intersection is not a common face.
We can thus construct the finest possible $\Sigma_b$ by enlarging
cones in $\tilde\Sigma_b$ (and throwing away subcones).

Since there are only finitely many triangulations of
$\partial^1\Nabla$, a simple approach to check the existence of
compatible fans is given by simply trying to construct a
compatible $\Sigma_b$ for each of them. Unfortunately, this
approach is computationally feasible only for the very simplest
cases (for $\dim \Nabla > 3$ the number of possible triangulations
is always finite, but almost always huge\footnote{If one wants to
enumerate all triangulations, the computational complexity is
largely reduced by only considering regular triangulations, i.e.
triangulations for which the resulting varieties are K\"ahler. One
can then calculate the secondary polytope instead of a brute force
enumeration. Although this reduces the complexity of the
enumeration from NP-hard to polynomial, the number of regular
triangulations is still huge in most cases.}).

If one puts some weak restrictions on the fibrations though, it is
possible to directly construct \emph{some} compatible
triangulation, if any exists. The restrictions used and the
methods for constructing the triangulations will be discussed in
section \ref{obssearch}.

\subsection{Multifibrations}
In this section $A$-fibered $B$-fibrations $X$ will be discussed.
The obvious interpretation of the above term is that $X$ is a
fibration, whose generic fiber is an object of type $B$, which
itself is a fibration with generic fiber $A$. One could also
define it as a space $X$, which is both an $A$- and a
$B$-fibration. I will use the term in a more restrictive way and
demand both. Phrased differently, I demand that the $B$- fibration
$X$ also is an $A$-fibration and the fibration structures are
compatible. In order to make the last statement precise, I will
introduce a third fibration as follows\footnote{The structures are
all very simple, but can be confusing on first sight. The latter
is reflected in the fact, that no good comprehensive diagrammatic
notation for multifibrations seems to exist.}.

\begin{defn}
An $A$-fibered $B$-fibration is a chain of morphisms
$$X \stackrel{\pi_1}{\longrightarrow} Y \stackrel{\pi_2}{\longrightarrow} Z$$
with the following properties:
\begin{itemize}
\item[(i)] $X \stackrel{\pi_1}{\longrightarrow} Y$ is an
$A$-fibration. $A$ is called the small generic fiber and $Y$ the
large base.
\item[(ii)] $Y \stackrel{\pi_2}{\longrightarrow} Z$ is a fibration
with generic fiber $F$. $Z$ is called the small base and $F$ the
intermediate base.
\item[(iii)] $X \stackrel{\pi_2\circ\pi_1}{\longrightarrow} Z$ is a $B$-fibration.
$B$ is called the generic large fiber.
\end{itemize}
\end{defn}
Note, that these conditions force the generic fibers of $X
\stackrel{\pi_2\circ\pi_1}{\longrightarrow} Z$ to be
$A$-fibrations over the intermediate base $F$.

In order to simplify the discussion, I will now specialize to the
case of elliptic K3 fibered Calabi-Yau hypersurfaces (either
threefolds or fourfolds) in toric varieties. With the above
notations, the small fiber $A=T^2$ is an elliptic curve, the large
fiber $B$ is K3, and the immediate base $F$ is $\PP^1$.

In complete analogy to the case of a single fibration, a
\textbf{toric elliptic K3 fibration} is the restriction to a
generic Calabi-Yau hypersurface of a \textbf{toric elliptic K3
prefibration}. The latter consists of a virtual prefibration
$$\Nabla_{ell} \subset \Nabla_{K3} \subset \Nabla$$
together with compatible base fans $\Sigma_{b, ell}, \Sigma_{b,
K3}$ and a compatible triangulation of $\partial^1\Nabla$. For the
prefibration, $A$ is $X_{\Delta_{ell}}$, $B$ is $X_{\Delta_{K3}}$
and $F$ is again $\PP^1$.

To give the notation intuitive meaning, we rename $\pi_1$ to
$\pi_e$ and define $\pi_K \defas \pi_2 \circ \pi_e$.

As in the case of a single fibration one can split\footnote{As
before, the splitting is not uniquely determined.} $N = N_e \oplus
\ZZ \oplus N_{b,K}$ with $N_K = N_e \oplus \ZZ$ and $N_{b,e} = \ZZ
\oplus N_{b,K}$ and obtain a commutative diagram
\[\begin{array}{ccccccccc}
 & & & & 0 & & 0 & & \\
 & & & & \downarrow & & \downarrow & & \\
0 & \rightarrow & N_e & \hookrightarrow & N_K &
\stackrel{\hat\pi_{e,K}}{\twoheadrightarrow} & \ZZ & \rightarrow & 0 \\
 & & || & & \downarrow & & \downarrow & & \\
0 & \rightarrow & N_e & \hookrightarrow & N &
\stackrel{\hat\pi_{e}}{\twoheadrightarrow}& N_{b,e} & \rightarrow & 0 \\
 & & & & \hat\pi_K\,\downarrow\,\phantom{\hat\pi_K} &
  & \phantom{\hat\pi_2}\,\downarrow\,\hat\pi_2 & & \\
 & & & & N_{b,K} & = & N_{b,K} & & \\
 & & & & \downarrow & & \downarrow & & \\
 & & & & 0 & & 0 & & \\
\end{array}\label{faserungskommutat}\]
with exact rows and columns.

\subsection{Algorithms for finding virtual fibrations}
\label{virtalgo} Given a collection of $D$-dimensional reflexive
polyhedra, one would like to find all toric Calabi-Yau fibration
structures carried by generic members of the corresponding
families of toric Calabi-Yau hypersurfaces. As a first step
towards this aim, one obviously has to find all virtual fibration
structures, i.e. reflexive subpolyhedra.

\subsubsection{Enumerating sublattices} \label{sublatalgo}
The most simple way to find reflexive subpolyhedra of a given
$D$-dimensional polyhedron $\Nabla \subset N_\RR$ uses the fact,
that any $d$-dimensional lattice subpolyhedron $\Nabla^{(f)}$
defines the $d$-dimensional lattice subspace $N^{(f)}_\RR$ spanned
by its vertices, which in turn are integer points of $\Nabla$.
Thus, the potential lattice subspaces can be enumerated via
$d$-tuples of integer points of $\Nabla$. These subspaces can then
be intersected with $\Nabla$ yielding a (not necessarily integer)
polyhedron $\Nabla^{(f)}$, which then can be checked for
reflexivity. The check for reflexivity can be performed as
follows:

Consider the lattice subspace $N^{(f)}_\RR$ spanned by integer
points $n_1, \ldots, n_d$ of $\Nabla$. We first
calculate\footnote{One can use Gaussian elimination to calculate
the kernel of $(n_1 \cdots n_d)$. This is a lattice subspace of
$M$, because $(n_1 \cdots n_d)$ is an integer matrix. Then one can
use the methods presented in appendix \ref{completetobase} to
calculate the bases (\ref{lattbasec1}) and (\ref{lattbasec2}). If
the kernel is not $(D-d)$-dimensional, the points are linearly
dependent and the result might thus be reflexive, but obviously
not $d$-dimensional.} a base
\[\left\{k_1,\ldots,k_{D-d},b_1,\ldots,b_d\right\}\label{lattbasec2}\]
 of $M$, where
\[\left\{k_1,\ldots,k_{D-d}\right\}\label{lattbasec1}\]
 is a base of
$N^{(f)} := (N^{(f)}_\RR)^\perp \cap M$. Let
$$K \defas \left(k_1 \cdots k_{D-d}\right),
\;c_N \defas \left(b_1 \cdots b_d\right)^t, \;B \defas
\left(\begin{array}{c}K^t\\c_N\end{array}\right)$$ and $\pi_M$ be
the last $d$ rows of $\left(B^{-1}\right)^t$. Note, that the
columns of $\pi_M^t$ form a base of the sublattice $N^{(f)}$.
Using this base and its dual base for $M^{(f)}$, $\pi_M$ is the
matrix of the projection $\pi_M:\,M\rightarrow M^{(f)}$. We have
$n_f \in N^{(f)}\,\Leftrightarrow\,K^t n_f = 0$ and the $n_f$ are
transformed to the base $\pi_M^t$ of the sublattice by the matrix
$c_N$.

Using $K^t$ we can immediately determine all elements of $N^{(f)}$
among the integer points of $\Nabla$. This information can be used
to avoid multiply checking the same subspace defined by different
tuples of points. In addition, it provides a first simple check
for reflexivity. If $\Nabla^{(f)}$ is to be reflexive at least its
vertices and the origin are integer points. Hence, we can
immediately dismiss the subspace if we do not find at least $d+2$
points.

Now let $\{m_i\}$ be the set of inner face normals of $\Nabla$,
i.e.\ the vertices of $\Delta = \Nabla^\star$. Obviously, $\{\pi_M
m_i\}$ generates $\Delta^{(f)} \defas (\Nabla^{(f)})^\star$ as a
convex set. From the inner point criterion for reflexivity, none
of the $\pi_M m_i$ can have integer length $> 1$ if $\Nabla^{(f)}$
is to be reflexive.

Assume, that all of the above tests are passed. The final test for
reflexivity can then be performed in either of two ways:

\begin{itemize}
\item One calculates the convex hull of $\{\pi_M m_i\}$ in terms
of its hypersurfaces. If (and only if) all inner face normals are
integer\footnote{Remember, that face normals are normalized to
have inner product -1 with the face.}, $\Delta^{(f)}$ is reflexive
and so is $\Nabla^{(f)}$.
\item Denote the convex hull of the integer points of $\Nabla^{(f)}$ by
$C$. $\Nabla^{(f)}$ is reflexive if and only if $C = \Nabla^{(f)}
= (\Delta^{(f)})^\star$. One calculates the inner face normals of
$C$, which must be integer for $\Nabla^{(f)}$ to be reflexive. In
this case, though, this is not sufficient. $C$ could be smaller
than $\Nabla^{(f)}$, in which case $C^\star$ is larger than
$\Delta^{(f)}$. If $\Nabla^{(f)}$ is reflexive, the vertices of
$C^\star$ are the vertices of $\Delta^{(f)}$ and thus are elements
of $\{\pi_M m_i\}$. If it is not, there must be some vertex of
$C^\star$ not contained in $\Delta^{(f)}$ and hence not an element
of $\{\pi_M m_i\}$.
\end{itemize}
The most expensive part in either of the two ways is calculating
the convex hull (the inner face normals) of a finite point set.
One should thus choose between the two ways depending on the size
of the corresponding point set.

Anyway, the check for reflexivity does not pose a serious
computational threat in any explicit example. The problem with the
algorithm is rather given by the number of subspaces to check,
which badly scales with the number $n$ of integer points in
$\Nabla$ approximately like $\binom{n}{d} \sim n^d$. For fourfold
polyhedra this poses a serious threat, since one easily finds
examples of polyhedra with $10^4$ - $10^5$ integer points.

The problem is obviously less serious, if one does not need to
enumerate all lattice subspaces, e.g. by fixing some subspace. I
actually used sublattice enumerations for finding reflexive
three-dimensinal subpolyhedra, which are superpolyhedra of a given
two-dimensional subpolyhedron. The number of subspaces one has to
check in such a setting for this task only scales linearly with
the number of integer points.

\subsubsection{Making use of classifications} \label{classialgo}
Instead of directly searching for reflexive subpolyhedra, one can
also use the dual picture and look for projections $\pi_M: \Delta
\twoheadrightarrow \Delta^{(f)}$. This was used by the authors of
\cite{searchk3fib}, who looked for reflexive faces of $\Delta$ and
projections to these faces. I did not follow this path, because it
further reduces the generality of the fibrations one finds. More
seriously, it inherently carries the risk of missing some of the
most interesting structures.

One cannot simply enumerate all possible projections except by
enumeration of subspaces as in the preceding section (which would
mean no gain), because the kernel of the projection map does not
have to generated by integer points of $\Delta$.

For cases, where a classification of the reflexive polyhedra in
$d$ dimensions is at hand, the projection approach nevertheless
provides an algorithm, which is almost\footnote{Almost, because it
does depend on the number of hypersurfaces of $\Nabla$, which is
not statistically independent.} independent of $n$. The main idea
is to search for projections $\pi_M: \Delta \twoheadrightarrow
\Delta^{(f)}$ for all possible reflexive images $\Delta^{(f)}$.

Let $(\Delta = \Nabla^\star, M)$ be as before and $(\Delta^{(f)},
M_f)$ a reflexive polyhedron of dimension $d$. We need to know,
whether there is a lattice projection $\pi_M: M \twoheadrightarrow
M_f$, which maps $\Delta$ to $\Delta^{(f)}$.

If there is such a map, it will map the integer points of $\Delta$
onto the integer points of $\Delta^{(f)}$. In particular, it will
map the vertices of $\Delta$ to integer points of $\Delta^{(f)}$
and due to linearity all vertices of $\Delta^{(f)}$ will be images
of vertices of $\Delta$. We thus choose $D$ linearly independent
vertices $m_1, \ldots, m_D$ of $\Delta$. Their images completely
determine $\pi_M$ by linearity. Hence, we can walk through all
maps $\{m_1, \ldots, m_D\} \rightarrow \{\mbox{integer points of
$\Delta^{(f)}$}\}$, reconstruct $\pi_M$ for each map and check
whether it fulfills the conditions. In detail, we check if
\begin{itemize}
\item $\pi_M$ is integer.
\item $\pi_M$ has maximal rank.
\item $\pi_M$ is surjective: There must be a primitive sublattice of
$M$ isomorphic to $M_f$. The inverse image of our base for $M_f$
under the isomorphism can be completed to a base for $M$. Using
this base, the matrix of $\pi_M$ is obtained by omitting rows from
the unit matrix. Hence, the rows of $\pi_M$ must be completeable
to a basis for $M$. The latter can be checked using algorithm
\ref{multicomplete}.
\item $\pi_M$ maps the vertices of $\Delta$ into $\Delta^{(f)}$.
\item one can find preimages of all vertices of $\Delta^{(f)}$
among the vertices of $\Delta$.
\end{itemize}

Of course, the number of mappings one has to check could be
significantly reduced by using known automorphisms of $\Delta$.
Since one usually does not know these beforehand and calculating
them is computationally expensive, I did not implement this.

Properties of members of the classification list, on the other
hand, are well known beforehand and can be used to optimize the
algorithm. In particular, one can
\begin{itemize}
\item[a)] use automorphisms of the $\Delta^{(f)}$ in order to
reduce the number of potential mappings and
\item[b)] use embeddings $\Delta^{(f)}_1 \hookrightarrow \Delta^{(f)}_2$
to search for projections to $\Delta^{(f)}_1$ while searching
those to $\Delta^{(f)}_2$.
\end{itemize}
Note though, that (a) and (b) partially exclude each other.

\section{Elliptically fibered toric K3 surfaces} \label{ellfibk3}
The basic duality between heterotic string and F-theory is between
F-theory compactified on an elliptically fibered K3 and the
heterotic string compactified on the generic fiber. This makes
elliptically fibered K3 surfaces an interesting object of study.

As mentioned in the introduction, the fiberwise extension of the
basic duality leads to a duality between F-theory on an elliptic
K3 fibered Calabi-Yau variety and the heterotic string on an
elliptic Calabi-Yau threefold.

The perturbative gauge group of the heterotic string theory dual
to F-theory on an elliptic K3 fibered Calabi-Yau variety is then
encoded in the generic K3 fiber and global monodromy arguments.
Thus, a good knowledge of elliptically fibered K3 surfaces will
also help in reading off properties of fibrations.

Elliptically fibered K3 surfaces are the easiest examples of toric
Calabi-Yau fibrations. In addition to the low dimensionality, the
codimension of the fiber is one. Hence, any virtual Calabi-Yau
prefibration gives rise to an elliptic fibration.

In the following sections I will often use properties of the
Picard lattice of a generic toric K3 surface, including its
intersection form. A full derivation of the formulae used to
calculate this lattice can be found in \cite{latticek3}. For the
reader's convenience, they are summarized in appendix
\ref{K3latticeformulae}.

\subsection{The class of the fiber}
The definition of F-theory on a Calabi-Yau manifold requires it to
be elliptically fibered, and for taking the F-theory limit one
needs a section of the fibration
$$\begin{array}{ccc}E & & \\ \downarrow & & \\ Y & \longrightarrow &
B.\end{array}$$
 The image of the section $s: B \rightarrow Y$ is an
effective divisor having intersection number 1 with the generic
fiber. Specializing to toric fibrations I will only consider
divisors of $Y$, which are sums of pullbacks of toric divisors or
irreducible components thereof. In other words, I consider divisor
classes of $Y$ corresponding to integer points of
$\partial^1\Nabla$ (either directly or as one of the irreducible
components of the toric divisor's intersection with the Calabi-Yau
hypersurface). Such divisors will be called
\textbf{semitoric}\footnote{For $Y$ a K3 manifold, a dense subset
of the parameter space of defining polynomials has enhanced Picard
group, i.e.\ $Pic(Y)$ is not generated by irreducible components
of the intersections of toric divisors with the K3 surface. For
higher dimensional $Y$ and generic (in the algebraic sense)
defining polynomials all divisor classes are semitoric.}.

In order to calculate the intersection number with the generic
fiber, one first needs to know the class of the generic fiber.
This is particularly simple in the case, where $Y$ a K3
surface\footnote{The following derivation may be easily
generalized to higher codimension by using multiple rational
functions and intersecting the corresponding divisors.}. Then, the
class of the generic fiber is itself a divisor of $Y$. The generic
fiber is the intersection of the K3 surface $Y$ with the generic
fiber of the toric prefibration
$$\begin{array}{ccc}X_{\Delta^{(f)}} & & \\ \downarrow & & \\ X_\Delta & \longrightarrow &
B,\end{array}$$
 i.e. the prefibration fiber over a generic point
in the torus $\CC^\star$ in the base $B = \PP^1$. The generic
point in the base is the set of zeroes of the rational function
$\chi^{m_b} - c \chi^0$, $c\in\CC^\star$ generic, on the base. Its
preimage is given by the set of zeroes of the pullback:
$$Z(\chi^{\hat\pi^t m_b} - c)$$
 Since $\hat\pi$ is surjective, $m_f \defas \hat\pi^t m_b$ is one of the two
primitive elements of $M$ perpendicular to $\Nabla^{(f)}$. The
divisor class $D_f$ is then readily calculated as
\begin{eqnarray}
D_f & = & Z(\chi^{m_f} - c) - div(\chi^{m_f} - c) \nonumber \\
 & = & \sum_{i,\langle m_f,n_i \rangle > 0} \langle m_f,n_i \rangle
D_i \label{firstfibform} \\
 & = & - \sum_{i,\langle m_f,n_i \rangle < 0} \langle
m_f,n_i \rangle D_i. \label{secondfibform}
\end{eqnarray}
 As usual, $D_i$ are the toric divisors of $X_\Delta$ corresponding
to primitive ray generators $n_i$. $div(f)$ denotes the principal
divisor of the rational function $f$. The last equality uses the
identity $\sum_i \langle m_f,n_i \rangle D_i = 0$ in
$A^1(X_\Delta)$ and reflects the fact, that the result is
independent of the choice of $m_b$.

\subsection{Gauge algebras} \label{pertgroupcalc}
Given an elliptic K3 surface in terms of a virtual elliptic
prefibration $\Nabla_{ell} \subset \Nabla_{K3}$ and a generic
defining polynomial $p$, we now want to read off the perturbative
gauge algebra of a dual heterotic string theory. As explained
e.g.\ in \cite{liecy3andf} and references therein, this is best
done in the type IIA theory compactified on the K3 surface. To
obtain F-theory, one first goes to the strong coupling limit
opening an effective additional dimension yielding M-theory. Then,
the F-theory limit is taken by shrinking all components of the
elliptic fiber to zero.

The last step is performed in two stages: First we shrink all
components of the fiber not meeting the chosen section of the
elliptic fibration\footnote{Taking the F-theory limit requires a
section of the fibration or a B-field in the base
\cite{candowithb1, candowithb2}. We will not discuss the latter
alternative.}. This leaves us with type IIA (or M-theory) on a
singular K3 surface. In the second step on shrinks the remaining
components to obtain F-theory. The gauge group then already
emerges in the type IIA theory\footnote{in the form of massless
twobranes wrapping the singularities. The gauge fields on these
branes interact according to the intersection pattern of the
exceptional divisors.}.

In order to read off the gauge algebra, we thus first have to
identify the divisor class of the section. The section is a
rational curve, and thus we have to look for effective divisors
$\tilde{D}_s$ of self-intersection\footnote{Recall, that the
adjunction formula forces any curve $c$ algebraically embedded
into a surface to have self-intersection $c^2 = 2 (g-1)$, where
$g$ is the genus of the curve.} $\tilde{D}_s^2 = -2$ and
intersection $D_f \cdot \tilde{D}_s = 1$ with the fiber.

From the latter condition and (\ref{firstfibform}),
(\ref{secondfibform}) we can deduce, that
$$\tilde{D}_s = \tilde{D}_{n_s} + \ldots,$$
where $\tilde{D}_{n_s}$ is an irreducible component of the toric
divisor corresponding to a vertex $n_s$ of $\Nabla_{ell}$ and
$(\ldots)$ is an effective semitoric divisor in the orthogonal
complement of $D_f$ (i.e. a sum of irreducible components of toric
divisors corresponding to points \emph{not} being vertices of
$\Nabla_{ell}$). Let us first assume that $\tilde{D}_s =
\tilde{D}_{n_s}$. Later it will be shown, that this assumption
does not restrict generality.

By explicit calculation it can easily be seen, that $\tilde{D}_s^2
= -2$ is already implied by $D_f \cdot \tilde{D}_s = 1$:

If $n_s$ lies in the interior of an edge of $\Nabla_{K3}$, all
irreducible components of $D_s \cap K3$ automatically have
self-intersection $-2$. This in particular treats the case, where
$D_s \cap K3$ splits into multiple irreducible components. In this
case it is also clear, that $n_s$ has precisely two neighboring
integer points on the common edge, which will be denoted by $n_u$
and $n_d$. The corresponding semitoric devisors having nonzero
intersection with $\tilde{D}_s$ will be denoted $\tilde{D}_u$ and
$\tilde{D}_d$. From (\ref{firstfibform}), (\ref{secondfibform})
and intersection number 1 with the fiber $\tilde{D}_f \defas D_f
\cap K3$ ($D_f \cdot \tilde{D}_s = 1$) we conclude $\tilde{D}_u
\cdot \tilde{D}_s = \tilde{D}_d \cdot \tilde{D}_s = 1$.

Now let $n_s$ be a vertex of $\Nabla_{K3}$. Then\footnote{The
following calculations are valid whenever this weaker condition is
fulfilled.}, $D_s \cap K3 = \tilde{D}_s$. Again using
(\ref{firstfibform}), (\ref{secondfibform}) and $D_f \cdot
\tilde{D}_s = 1$ we can conclude that as before $n_s$ must have
exactly one neighbor $n_u$ above and $n_d$ below $\Nabla_{ell}$ on
a common edge. In addition, the dual edges to these edges must
have integer length $1$ and both $n_u$ and $n_d$ must have integer
distance $1$ from the plane of $\Nabla_{ell}$ (as measured by
$m_f$).

Denote by $e_1$ and $e_2$ the integer points neighboring $n_s$ in
$\Nabla_{ell}$ (which might either lie on a surface or an edge of
$\Nabla_{K3}$). Both $\{n_s, e_1\}$ and $\{n_s, e_2\}$ must be
bases of $N_{ell} \cong \ZZ^2$. By choice of base we can thus
write
$$n_s = \threevec{1}{0}{0},\;e_1 = \threevec{0}{1}{0},\;e_2 =
\threevec{\alpha}{-1}{0},\; n_d = \threevec{1}{0}{-1},\; n_u =
\threevec{1-\beta}{\gamma}{1}.$$
 One calculates the (not necessarily pairwise different) face
 normals
$$m^1_d = \threevec{-1}{-1}{0},\; m^2_d =
\threevec{-1}{1-\alpha}{0},\; m^1_u =
\threevec{-1}{-1}{\gamma-\beta}\;\mbox{and}\; m^2_u =
\threevec{-1}{1-\alpha}{\gamma(\alpha-1)-\beta}.$$
 Since $n_s$ is a vertex of $\Nabla_{ell}$, we have $\alpha \le
1$. The length of the edge $\theta_d$ dual to $\overline{n_s n_d}$
is $$l(\theta_d) = |m^1_d - m^2_d| = |2-\alpha| \stackrel{!}{=}
1\qquad \Rightarrow \qquad \alpha = 1,$$ which also ensures
$l(\theta_u)=1$. We must have $\langle m^{1/2}_d, n_u\rangle \ge
-1$ and hence $\beta \ge \gamma$ and $\beta \ge 0$.

 Denote by $E_1,E_2,D_d,D_u$ the intersections
of the K3 surface with the toric divisors corresponding to
$e_1,e_2,n_d,n_u$. For the self-intersection of $\tilde{D}_s$ we
obtain
\begin{eqnarray*}
\tilde{D}_s^2 & = & \langle m^2_d, e_1\rangle \tilde{D}_s E_1 +
\langle m^2_d, e_2\rangle \tilde{D}_s E_2 + \langle m^2_d,
n_d\rangle \tilde{D}_s D_d + \langle m^2_d, n_u\rangle \tilde{D}_s
D_u \\
& = & \langle m^2_d, e_1\rangle |m^1_u-m^1_d| + \langle m^2_d,
e_2\rangle |m^2_u-m^2_d| + \langle m^2_d, n_d\rangle + \langle
m^2_d, n_u\rangle \\
& = & 0 - 1\cdot \beta -1 + (\beta-1) = -2.
\end{eqnarray*}

The divisor $\tilde{D}_f$ of the fiber is the class of an elliptic
curve. It thus has self-intersection $\tilde{D}_f^2 = 0$, which
can also be read off using both (\ref{firstfibform}) and
(\ref{secondfibform}). Therefore, the system $\tilde{D}_f$ and
$\tilde{D}_s$ span a hyperbolic lattice $H = \langle \tilde{D}_f,
\tilde{D}_s + \tilde{D}_f \rangle_\ZZ$. Since $H$ is self-dual,
the Picard lattice splits as
$$Pic(K3) = H \oplus H^\perp.$$
This follows from the corresponding splitting of $H^2(K3,\ZZ)$
\cite[\S 1]{nikulinlemma}\footnote{Since the reference is not
available online, you might want to look up the relevant lemma in
\cite[Lemma 4.6]{latticek3}.}. $H^\perp \subset Pic(K3)$ must be
negative definite, because it is contained in the intersection of
$H^2(K3,\ZZ)$ with the orthogonal complement of the positive
definite threeplane in $H^2(K3,\RR)$ spanned by the real and
imaginary parts of the complex structure and $\tilde{D}_s + 2
\tilde{D}_f$.

\begin{rem}
We can now see, why our initial assumption $\tilde{D}_s =
\tilde{D}_{n_s}$ does not mean a loss of generality. Any other
choice would have been of the form
$$\tilde{D}'_s = \tilde{D}_s + \alpha \tilde{D}_f + h,\quad \alpha\in\ZZ, h\in H^\perp.$$
Now ${\left.\tilde{D}'_s\right.}^2 = -2 \;\Rightarrow\; \alpha =
-h^2$. Let $\{h_i; i\in I \}$ be a basis of $H^\perp$. The change
of basis $\tilde{D}_s \mapsto \tilde{D}'_s$, $\tilde{D}_f \mapsto
\tilde{D}_f$ and $h_i \mapsto h'_i \defas h_i - (h \cdot h_i)
\tilde{D}_f$ is an automorphism of $Pic(K3)$. As $Pic(K3)$ is a
primitive sublattice of $H^2(K3,\ZZ)$, one can use the results of
\cite{nikulinlemma} to lift this automorphism to $H^2(K3,\ZZ)$
extending the identity on $Pic(K3)^\perp \subset H^2(K3,\ZZ)$. Due
to the results of \cite{latticek3}, $Pic(K3)^\perp$ always
contains a hyperbolic sublattice denoted by $\tilde{H}$ in
\cite{latticek3}. If necessary, the lifted automorphism can thus
be made orientation-preserving by exchanging the generators of
$\tilde{H}$. Being orientation-preserving, the automorphism is
induced by a diffeomorphism of the K3 surface \cite{k3diffeo1,
k3diffeo2, k3diffeo3}.

By similar arguments, any two choices of hyperbolic sublattice are
connected by a diffeomorphism of the K3 surface. The important
point is that in our case the plane of the complex structure
(resp.\ the Picard lattice) is preserved. Note, that one
\emph{can} have inequivalent (i.e. different modulo automorphisms
of $Pic(K3)$) embeddings $H \hookrightarrow Pic(K3)$ when choosing
a different $n_s$.
\end{rem}

Two sublattices of $H^\perp \subset Pic(K3)$ can be easily
extracted, namely the lattices $\Gamma_u$ and $\Gamma_d$ spanned
by semitoric hypersurface divisors corresponding to points above
(below) $\Nabla_{ell}$ apart from $\tilde{D}_u$ ($\tilde{D}_d$).

For dimensionality reasons the space of relations between
semitoric divisors is precisely the space of relations between the
corresponding divisors of the ambient toric variety, which is just
$M \cong M_{ell} \oplus \ZZ m_f$. $m \in M$ corresponds to the
relation
\begin{eqnarray*}0 & \equiv & \sum_{n \in \partial\Nabla_{K3}
\cap N} \langle m, n\rangle D_n \cap K3 \\
& = & \sum_{n \in \partial\Nabla_{K3} \cap N} \langle m, n\rangle
\sum \tilde{D}_n.
\end{eqnarray*}
The only relations involving only divisors corresponding to points
outside the elliptic plane are those given by multiples of $m_f$
and thus always contain $\tilde{D}_u$ and $\tilde{D}_d$ with
nonzero coefficients.

Hence, the above semitoric divisors are a base of $\Gamma_u \oplus
\Gamma_d$. By negative definiteness they must be rational curves.

So far, I have only talked about smooth elliptically fibered K3
surfaces. We now shrink to zero all the rational curves in
$\Gamma_u \oplus \Gamma_d$. If there is anything to shrink, our K3
surface will develop (one or) two pointlike singularities located
at the points $0$ and $\infty$ over the base $\PP^1$. The ADE
classification of these singularities is then nothing else than
the intersection form on exceptional divisors of the blowup (i.e.
the original smooth K3 surface) and hence coded in the
intersection matrix on $\Gamma_u$ and $\Gamma_d$, respectively. By
determining the ADE type, one also reads off the gauge algebras
corresponding to these singularities\footnote{The intersection
form is minus the Cartan matrix, i.e. (all roots have equal
length) is proportional to the Killing form. The latter statement
directly extends to $\Gamma_u \oplus \Gamma_d \oplus H$: Here the
intersection form is an invariant bilinear form on the Cartan
subalgebra of the corresponding extended untwisted affine
Kac-Moody algebra.}.

\begin{rem} If none of the toric divisors corresponding to points
in the upper and lower half of the K3 polyhedron splits into
multiple semitoric divisors, the Dynkin diagrams may be read off
directly from the polyhedron's edge diagram as observed in
\cite{groupvisibleinpoly}. The only detail not directly obvious
from the edge diagram is that divisors corresponding to vertices
of $\Nabla_{K3}$ have the correct self-intersections, which
follows from negative definiteness.

In the split case, one can still read off the Dynkin diagram, if
one splits the points and edges corresponding to split toric
divisors and the intersections between them. Remember, that in
contrast to the nonsplit case, the points above and below the
section point must not be omitted, but only one irreducible
component of each of them.
\end{rem}

\begin{rem}
In simple examples (as those mainly studied in
\cite{k3chowformula}), one has $\Gamma_u \oplus \Gamma_d = H^\perp
\subset Pic(K3)$. This, of course, is not always the case. In
particular, two things can happen:
\begin{itemize}
\item[i)]
There could be rational curves apart from those in $\Gamma_u
\oplus \Gamma_d$ which one can shrink to zero to obtain enhanced
gauge symmetry. Shrinking to zero divisors corresponding to
vertices of $\Nabla_{ell}$ leads to a globally singular
``elliptic'' fibration, since these divisors do meet the generic
fiber. Divisors corresponding to points on edges of both
$\Nabla_{ell}$ and $\Nabla_{K3}$, on the other hand, do not meet
the generic fiber. Since they \emph{do} meet the generic K3
hypersurface, shrinking them to zero introduces singularities over
points in $\TTT \subset \PP^1$.

Rational curves corrsponding to such points intersect neither
those in $\Gamma_u$ and $\Gamma_d$ nor those corresponding to
points on different edges of $\Nabla_{ell}$. Hence, one might only
obtain additional gauge group factors of type $A_n$.
\item[ii)] The lattice $\Gamma_u \oplus \Gamma_d$ is not primitive.
This happens, whenever the integer points in the elliptic plane
not lying on hypersurfaces of $\Nabla_{K3}$ together with $n_u$
and $n_d$ do not generate $N$: Precisely in this case there is a
primitive relation $m\in M$ between a nonprimitive element of the
lattice spanned by the divisors corresponding to the above points
and an element of the obvious lift of $\Gamma_u \oplus \Gamma_d$.

The occurence of additional $A_n$ factors as described in (i)
requires a common edge of $\Nabla_{ell}$ and $\Nabla_{K3}$. The
integer points lying on this edge provide a base for $N_{ell}$.
Therefore, both phenomena cannot occur simultaneously.

This is not a problem, but rather an interesting feature. It is
related to the fact, that $\Gamma_u \oplus \Gamma_d$ are the root
lattice of the corresponding gauge algebra, which often only is a
sublattice of the weight lattice. To illustrate the point,
consider the ``heterotic duality'' polyhedron (number 4319 in
\cite{frk3fiblist}) with vertices
$$n_{E_1} = \threevec{1}{0}{0},\,n_F=\threevec{0}{1}{0},\,
n_{C_0}=\threevec{3}{4}{6}\,\mbox{and}\;n_{C_{12}}=\threevec{-9}{-8}{-6}.$$
This polyhedron allows two elliptic fibrations (numbers 13277 and
13278 in \cite{frk3fiblist}) with section corresponding to the
planes with normals
$$\tilde{m}_f=\threevec{0}{0}{1}\:\mbox{and}\:m_f=\threevec{0}{1}{-1}.$$
While the former leads to gauge group $E_8 \times E_8$, the latter
leads to gauge algebra $D_{16}$.

Apart from the vertices, the integer points on edges of
$\Nabla_{K3}$ are
$$n_{S}=\threevec{-3}{-2}{-2},\,n_{d}=\threevec{-6}{-5}{-4},\,
n_{E_2}=\threevec{1}{2}{2},\,n_{A_1}=\threevec{2}{2}{3},\,
n_{A_2}=\threevec{2}{3}{4},$$
$$n_{C_{13}}=\threevec{-4}{-4}{-3}\quad \mbox{and}\quad
n_{C_i}=\threevec{3-i}{4-i}{6-i},i=1\ldots,11.$$
 $n_{E_1},n_{E_2}$ and $n_S$ are the vertices of the fiber
polyhedron. The divisor $S$ is a section (so is, by symmetry,
$E_2$) and, together with $F$, spans our hyperbolic lattice $H$.
With notations as before, $n_u = n_F$ and $\Gamma_u = \{ 0 \}$. A
base for $\Gamma_d$ is given by $A_1, A_2, C_0,\ldots C_{13}$. The
corresponding points, together with the edge segments between
them, form the Dynkin diagram of $D_{16}$. Simply counting
dimensions, one might think that $Pic(K3) = H \oplus \Gamma_d$.
This is not true, because $n_{E_1},n_{E_2},n_u,n_d$ and $n_S$ only
generate an index two sublattice of $N$. Thus, one needs one
additional generator for $H^\perp$. A simple calculation
evaluating the relations in $M$ shows that one can take
$$E_1 - 2 E_2 \equiv \frac{1}{2} \left( 5 A_1 + 8 A_2 + \sum_{i=0}^{13} (12-i) C_i \right),$$
which completes $H^\perp$ to the weight lattice of
$spin(32)/\ZZ_2$.

Note, that one can also derive the qualitative result without
doing any calculations at all: Considering the fibration given by
$\tilde{m}_f$ with gauge group $E_8 \times E_8$, one knows that
$Pic(K3)$ must be self-dual and so must then be our $H^\perp$.
Since there are only two even self-dual negative definite lattices
in 16 dimensions, it must be the weight lattice of
$spin(32)/\ZZ_2$.

\item[iii)] If one does not wish to ignore additional gauge
algebra summands as discussed in (i), the additional summand can
easily be calculated. The only thing one has to remember is to
omit semitoric divisors corresponding to points on common edges of
$\Nabla_{K3}$ and $\Nabla_{ell}$, which are neighbors of the
chosen section. For the additional $A_n$ summands, a situation as
described before for nontoric sections (omitting only one
irreducible component of a toric divisor) cannot occur: For
nontoric section $n_s$ must lie on a common split edge of
$\Nabla_{K3}$ with $n_u$ and $n_d$ and would thus need to be a
vertex in order to also lie on a common edge of $\Nabla_{K3}$ and
$\Nabla_{ell}$. This, in turn, contradicts the assumption of the
section to be nontoric. Arguments similar to the discussion of
$\Gamma_u \oplus \Gamma_d$ show, that one indeed obtains a
sublattice $\Gamma_u \oplus \Gamma_d \oplus A$, where $A$ denotes
the additional gauge algebra summand as directly read off from the
polyhedron. In (ii) I argued, that $\Gamma_u \oplus \Gamma_d$ must
be a primitive sublattice of $H^\perp$. This is \emph{not} true
for the full sum $\Gamma_u \oplus \Gamma_d \oplus A$. Primitivity
can fail if we cannot find a base for $N_{ell}$ consisting of
vertices of $N_{ell}$ only.

\end{itemize}
\end{rem}
\begin{table}[hbt]
\begin{center}
\begin{tabular}{|l|r|}
\hline
 \multicolumn{2}{|c|}{\textbf{Elliptically fibered toric K3 surfaces
(overview)}} \\ \hline\hline
 reflexive polyhedra & 4,319 \\
 elliptic fibrations & 13,278 \\
 \phantom{x} with section & 12,060 \\
 \phantom{x} \phantom{x} nontoric section only & 729 \\
 \phantom{x} nonabelian gauge algebra & 11,890 \\
 \phantom{x} non-simply laceable & 4,505 \\
 \phantom{x} \phantom{x} nontoric section & 376 \\
 \hline
\end{tabular}
\parbox{12cm}{\caption[Elliptically fibered toric K3 surfaces
(overview)]{{\bf Elliptically fibered toric K3 surfaces.\ }{\em
For elliptically fibered K3 surfaces, all virtual elliptic
fibrations are fibrations. The numbers in the table refer to
virtual elliptic fibrations after identification via automorphisms
of the K3 polyhedron. ``Non-simply laceable'' means, that
non-simply laced gauge algebras might emerge when the elliptic K3
is the generic fiber of an elliptic K3 fibration (c.f.\ section
\ref{monodromy1}).}}\label{k3ovv}}
\end{center}
\end{table}

I have searched all 4319 reflexive polyhedra from
\cite{kreuzskarkek3list} corresponding to toric K3 hypersurfaces
for toric elliptic fibrations. For all fibration structures, I
checked for toric and nontoric sections and the corresponding
gauge algebras by calculation and classification of the
intersection matrices of $\Gamma_u$ and $\Gamma_d$. Lists
containing the results can be found at \cite{frk3fiblist}. An
overview is given in table \ref{k3ovv}.

\subsection{Monodromy} \label{monodromy1}
\begin{figure}[htb]
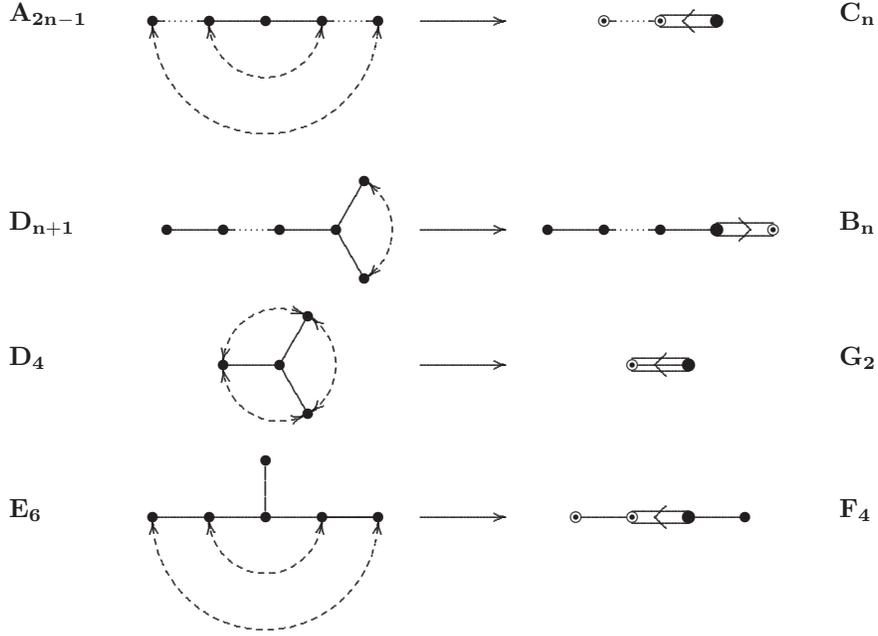

\setlength{\unitlength}{0.75cm}\begin{tabular}{l@{\hspace{0.5cm}}c@{}c@{}c@{\hspace{0.5cm}}l}
$\mathbf{A_{2n-1}}$ & \mbox{\beginpicture
 \setcoordinatesystem units <\unitlength,\unitlength>
 \setplotarea x from 0 to 5, y from -2.5 to 0.5
 \put {\circle*{0.2}} [Bl] at 0.5 0
 \put {\circle*{0.2}} [Bl] at 1.5 0
 \put {\circle*{0.2}} [Bl] at 2.5 0
 \put {\circle*{0.2}} [Bl] at 3.5 0
 \put {\circle*{0.2}} [Bl] at 4.5 0
 \plot 1.3 0 3.7 0 /
 \plot 0.5 0 0.7 0 /
 \plot 4.3 0 4.5 0 /
 \setdots <0.1\unitlength>
 \plot 0.5 0 1.5 0 /
 \plot 3.5 0 4.5 0 /
 \setdashes <0.1\unitlength>
 \circulararc -180 degrees from 4.5 0 center at 2.5 0
 \circulararc -180 degrees from 3.5 0 center at 2.5 0
 \setsolid
 \arrow <0.2\unitlength> [0.25,0.75] from 0.51 -0.2 to 0.5 -0.1
 \arrow <0.2\unitlength> [0.25,0.75] from 1.52 -0.2 to 1.5 -0.1
 \arrow <0.2\unitlength> [0.25,0.75] from 3.48 -0.2 to 3.5 -0.1
 \arrow <0.2\unitlength> [0.25,0.75] from 4.49 -0.2 to 4.5 -0.1
\endpicture} &
\mbox{\beginpicture
 \setcoordinatesystem units <\unitlength,\unitlength>
 \setplotarea x from 0 to 2, y from -0.5 to 0.5
 \arrow <0.2\unitlength> [0.25,0.75] from 0.25 0 to 1.75 0
\endpicture} &
\mbox{\beginpicture
 \setcoordinatesystem units <\unitlength,\unitlength>
 \setplotarea x from 0 to 3, y from -2.5 to 0.5
 \put {\circle{0.2}} [Bl] at 0.5 0
 \put {\circle*{0.1}} [Bl] at 0.5 0
 \put {\circle{0.2}} [Bl] at 1.5 0
 \put {\circle*{0.1}} [Bl] at 1.5 0
 \put {\circle*{0.22}} [Bl] at 2.5 0
 \plot 1.3 0 1.39 0 /
 \plot 0.61 0 0.7 0 /
 \plot 1.5 0.11 2.5 0.11 /
 \plot 1.5 -0.11 2.5 -0.11 /
 \plot 2.1 0.2 1.9 0 2.1 -0.2 /
 \setdots <0.1\unitlength>
 \plot 0.61 0 1.39 0 /
\endpicture} &
$\mathbf{C_n}$ \\
$\mathbf{D_{n+1}}$ & \mbox{\beginpicture
 \setcoordinatesystem units <\unitlength,\unitlength>
 \setplotarea x from 0 to 4.5, y from -1.2 to 1.2
 \put {\circle*{0.2}} [Bl] at 0.5 0
 \put {\circle*{0.2}} [Bl] at 1.5 0
 \put {\circle*{0.2}} [Bl] at 2.5 0
 \put {\circle*{0.2}} [Bl] at 3.5 0
 \put {\circle*{0.2}} [Bl] at 4 0.866
 \put {\circle*{0.2}} [Bl] at 4 -0.866
 \plot 0.5 0 1.7 0 /
 \plot 2.3 0 3.5 0 /
 \plot 3.5 0 4 0.866 /
 \plot 3.5 0 4 -0.866 /
 \setdots <0.1\unitlength>
 \plot 1.5 0 2.5 0 /
 \setdashes <0.1\unitlength>
 \circulararc -120 degrees from 4 0.866 center at 3.5 0
 \setsolid
 \arrow <0.2\unitlength> [0.25,0.75] from 4.15 0.76 to 4.09 0.82
 \arrow <0.2\unitlength> [0.25,0.75] from 4.15 -0.76 to 4.09 -0.82
\endpicture} &
\mbox{\beginpicture
 \setcoordinatesystem units <\unitlength,\unitlength>
 \setplotarea x from 0 to 2, y from -0.5 to 0.5
 \arrow <0.2\unitlength> [0.25,0.75] from 0.25 0 to 1.75 0
\endpicture} &
\mbox{\beginpicture
 \setcoordinatesystem units <\unitlength,\unitlength>
 \setplotarea x from 0 to 5, y from -1.2 to 1.2
 \put {\circle*{0.2}} [Bl] at 0.5 0
 \put {\circle*{0.2}} [Bl] at 1.5 0
 \put {\circle*{0.2}} [Bl] at 2.5 0
 \put {\circle*{0.22}} [Bl] at 3.5 0
 \put {\circle{0.2}} [Bl] at 4.5 0
 \put {\circle*{0.1}} [Bl] at 4.5 0
 \plot 0.5 0 1.7 0 /
 \plot 2.3 0 3.5 0 /
 \plot 3.5 0.11 4.5 0.11 /
 \plot 3.5 -0.11 4.5 -0.11 /
 \plot 3.9 0.2 4.1 0 3.9 -0.2 /
 \setdots <0.1\unitlength>
 \plot 1.5 0 2.5 0 /
\endpicture} &
$\mathbf{B_n}$ \\
$\mathbf{D_4}$ & \mbox{\beginpicture
 \setcoordinatesystem units <\unitlength,\unitlength>
 \setplotarea x from 0 to 2.5, y from -1.2 to 1.2
 \put {\circle*{0.2}} [Bl] at 0.5 0
 \put {\circle*{0.2}} [Bl] at 1.5 0
 \put {\circle*{0.2}} [Bl] at 2 0.866
 \put {\circle*{0.2}} [Bl] at 2 -0.866
 \plot 0.5 0 1.5 0 /
 \plot 1.5 0 2 0.866 /
 \plot 1.5 0 2 -0.866 /
 \setdashes <0.1\unitlength>
 \circulararc -360 degrees from 0.5 0.0 center at 1.5 0
 \setsolid
 \arrow <0.2\unitlength> [0.25,0.75] from 2.15 0.76 to 2.09 0.82
 \arrow <0.2\unitlength> [0.25,0.75] from 2.15 -0.76 to 2.09 -0.82
 \arrow <0.2\unitlength> [0.25,0.75] from 0.513 0.15 to 0.503 0.1
 \arrow <0.2\unitlength> [0.25,0.75] from 0.513 -0.15 to 0.503 -0.1
 \arrow <0.2\unitlength> [0.25,0.75] from 1.85 0.932 to 1.91 0.912
 \arrow <0.2\unitlength> [0.25,0.75] from 1.85 -0.932 to 1.91 -0.912
\endpicture} &
\mbox{\beginpicture
 \setcoordinatesystem units <\unitlength,\unitlength>
 \setplotarea x from 0 to 2, y from -0.5 to 0.5
 \arrow <0.2\unitlength> [0.25,0.75] from 0.25 0 to 1.75 0
\endpicture} &
\mbox{\beginpicture
 \setcoordinatesystem units <\unitlength,\unitlength>
 \setplotarea x from 0 to 2, y from -0.5 to 0.5
 \put {\circle{0.2}} [Bl] at 0.5 0
 \put {\circle*{0.1}} [Bl] at 0.5 0
 \put {\circle*{0.22}} [Bl] at 1.5 0
 \plot 0.5 0.11 1.5 0.11 /
 \plot 0.5 -0.11 1.5 -0.11 /
 \plot 0.61 0 1.4 0 /
 \plot 1.1 0.2 0.9 0 1.1 -0.2 /
\endpicture} &
$\mathbf{G_2}$ \\
$\mathbf{E_6}$ & \mbox{\beginpicture
 \setcoordinatesystem units <\unitlength,\unitlength>
 \setplotarea x from 0 to 5, y from -2.5 to 1.5
 \put {\circle*{0.2}} [Bl] at 0.5 0
 \put {\circle*{0.2}} [Bl] at 1.5 0
 \put {\circle*{0.2}} [Bl] at 2.5 0
 \put {\circle*{0.2}} [Bl] at 2.5 1
 \put {\circle*{0.2}} [Bl] at 3.5 0
 \put {\circle*{0.2}} [Bl] at 4.5 0
 \plot 0.5 0 4.5 0 /
 \plot 2.5 0 2.5 1 /
 \plot 3.5 0 4.5 0 /
 \setdashes <0.1\unitlength>
 \circulararc -180 degrees from 4.5 0 center at 2.5 0
 \circulararc -180 degrees from 3.5 0 center at 2.5 0
 \setsolid
 \arrow <0.2\unitlength> [0.25,0.75] from 0.51 -0.2 to 0.5 -0.1
 \arrow <0.2\unitlength> [0.25,0.75] from 1.52 -0.2 to 1.5 -0.1
 \arrow <0.2\unitlength> [0.25,0.75] from 3.48 -0.2 to 3.5 -0.1
 \arrow <0.2\unitlength> [0.25,0.75] from 4.49 -0.2 to 4.5 -0.1
\endpicture} &
\mbox{\beginpicture
 \setcoordinatesystem units <\unitlength,\unitlength>
 \setplotarea x from 0 to 2, y from -0.5 to 0.5
 \arrow <0.2\unitlength> [0.25,0.75] from 0.25 0 to 1.75 0
\endpicture} &
\mbox{\beginpicture
 \setcoordinatesystem units <\unitlength,\unitlength>
 \setplotarea x from 0 to 4, y from -2.5 to 0.5
 \put {\circle{0.2}} [Bl] at 0.5 0
 \put {\circle*{0.1}} [Bl] at 0.5 0
 \put {\circle{0.2}} [Bl] at 1.5 0
 \put {\circle*{0.1}} [Bl] at 1.5 0
 \put {\circle*{0.22}} [Bl] at 2.5 0
 \put {\circle*{0.20}} [Bl] at 3.5 0
 \plot 0.61 0 1.39 0 /
 \plot 1.5 0.11 2.5 0.11 /
 \plot 1.5 -0.11 2.5 -0.11 /
 \plot 2.5 0 3.5 0 /
 \plot 2.1 0.2 1.9 0 2.1 -0.2 /
\endpicture} &
$\mathbf{F_4}$
\end{tabular}
\begin{center}
\parbox{11cm}{\caption[Outer automorphisms of Lie algebras]{{\bf Outer automorphisms
of Lie algebras.\ }{\em The dashed arrows connect roots
interchanged by an outer automorphism. Note, that Cartan matrix
and intersection form are related by a simple switch of sign only
when all roots have equal length (i.e. on the left hand side). The
direction of the arrows on the right hand side may easily be
deduced by using Serre relations on Chevalley generators of the
invariant subalgebra. The problematic case of outer automorphisms
of $A_{2n}$ \cite{liecy3andf} cannot occur, because a partial
split toric edge diagram can only be connected when including a
vertex point (which is nonsplit).}}\label{lieautom}}\end{center}
\end{figure}

The main idea for extending the $Het \leftrightarrow F$ duality to
higher dimension is to use elliptic K3 fibrations and fiberwise
duality. If one demands purely geometric phases, the F-theory
compactification space in fact must be a K3 fibration
\cite{needk3fordual}. A phenomenon typical to fibrations can
modify the gauge group emerging in the pure K3 case, namely
monodromy. Following a closed path in the base of the K3 fibration
induces an automorphism of the K3 fiber. This induces an
automorphism of the Picard lattice, which does not need to be
trivial. The gauge group then only emerges as the monodromy
invariant part of the original Lie algebra. The relevant
automorphisms of Lie algebras are depicted in figure
\ref{lieautom}.

When studying toric eliptic K3 fibrations, one is interested in a
simple criterion for the occurrence of such identifications. As
the assumptions made in \cite{k3chowformula} are not always valid,
the simple criterion given there using only properties of
$\Nabla_{K3}$ will not suffice. We will see, though, that it can
easily be extended to full generality by taking into account
simple properties of the polyhedron $\Nabla_{CY}$ corresponding to
our elliptic K3-fibered toric Calabi-Yau hypersurface. Consider a
toric K3 fibration
$$\begin{array}{cccc}
K3 \subset X_{\Delta_{K3}} & & \\ \downarrow & & \\
CY \subset X_{\Delta_{CY}} & \longrightarrow & B\end{array}$$
 with virtual prefibration $\Nabla_{K3} \subset \Nabla_{CY}$.

The toric divisors of the generic prefibration fiber are just the
intersections of the fiber with the toric divisors of
$X_{\Delta_{CY}}$ corresponding to integer points in the K3
subpolyhedron $\Nabla_{K3}$. Hence, they cannot be interchanged
when moving around in the prefibration base. Obviously, the same
argument prevents toric divisors in the K3 fiber from being
permuted. We thus only have to find out, whether the nontoric
irreducible components of a given toric divisor are interchanged
by monodromy. First assume (as was done in \cite{k3chowformula})
that the corresponding divisor of $X_{\Delta_{CY}}$ is
irreducible. Assume furthermore, that the irreducible components
of the fiber divisor split into multiple orbits of the monodromy
group acting on them. For each orbit, we obtain an irreducible
component of the divisor of $X_{\Delta_{CY}}$ by moving around an
irreducible component of the fiber divisor over the base. We thus
obtain multiple irreducible components of the divisor of
$X_{\Delta_{CY}}$ in contradiction to the assumption.

Now assume, that the toric divisor of $X_{\Delta_{CY}}$ also
splits into multiple irreducible components. We will see, that no
identifications by monodromy occur in this case. Denote by $n_d
\in \Nabla_{K3} \subset \Nabla_{CY}$ the point corresponding to
our reducible toric divisor, which must lie on an edge
$\theta^\star_{K3}$ of $\Nabla_{K3}$ and in the interior of a
codimension 2 face $\theta^\star_{CY} \supset \theta^\star_{K3}$
of $\Nabla_{CY}$. Then the preimage of the dual edge $\theta_{K3}$
of $\Delta_{K3}$ under $\pi_M$ must precisely be the edge
$\theta_{CY}$ of $\Delta_{CY}$ dual to $\theta^\star_{CY}$. Hence,
the toric divisor of the K3 fiber splits into exactly equally many
irreducible components as the divisor of the total Calabi-Yau
variety, showing that no identifications occur.

The same argument applies to the section. A given nontoric section
of the generic elliptic K3 fiber of a toric elliptic K3 fibration
patches together to yield a section (in contrast to a
multisection) of the total elliptic fibration precisely when the
corresponding point lies in the interior of a codimension two face
of $\Nabla_{CY}$.

\relax

\subsection{A Note on Generality}
One may wonder whether we have lost elliptic fibrations by
requiring them to be toric, i.e.\ to be induced by a toric
prefibration. This is most certainly the case. A more general
approach would be to search for possible classes of the generic
fiber, i.e. to search for a pencil of effective divisors with
vanishing self-intersection. Finding such divisors is technically
difficult since the equation one wants to solve is quadratic
rather than linear.

In many cases without toric elliptic fibration, one can at least
show that a more general elliptic fibration does not exists,
either. This is clear for the cases with Picard number 1. In these
cases the unique divisor class has positive self-intersection.
Even when we have divisor classes with both positive and negative
self-intersection, often no divisor class with vanishing
self-intersection exists. For an easy example, consider the
reflexive polyhedron (no. 8 in the list at \cite{frk3fiblist})
with vertices
$$\threevec{1}{0}{0},
 \threevec{-1}{0}{0},
 \threevec{0}{1}{0},
 \threevec{0}{0}{1}\quad\mbox{and}\quad
 \threevec{2}{-1}{-1}.$$
The generic hypersurface has Picard number 2 and using the
formulae in appendix \ref{K3latticeformulae} one calculates the
intersection matrix to be
$$\left(\begin{matrix}
-2 & 1 \\
1 & 2
\end{matrix}\right).$$
I make the ansatz $E = x D_1 + y D_2$ for the class of an elliptic
fiber with $x,y \in \ZZ$. One then calculates
$$-2 x^2 + 2 y^2 + 2 xy \stackrel{!}{=} 0 \quad\Rightarrow\quad
 x_{1,2} = y \frac{1 \pm \sqrt{5}}{2},$$
which has no solution in $\QQ^2$.

\section{Fourfolds}
As stated before, one obtains models with four flat space-time
dimensions by compactification of the heterotic string on a
Calabi-Yau threefold with vector bundle. F-theory, on the other
hand, has to be compactified on an elliptic Calabi-Yau fourfold.

For an overview of different ways to obtain explicit examples of
Calabi-Yau fourfolds as well as general properties and relations
between invariants of such varieties, the interested reader is
referred to \cite{klyr}. In order to use the methods presented in
the preceding section, the focus of interest in this paper is
naturally given by hypersurfaces in five-dimensional toric
varieties.

Due to the lack of a complete classification of reflexive
Polyhedra in five dimensions one has to resort to constructible
subclasses of polyhedra in order to obtain explicit examples to
study. With the focus of obtaining families of varieties well
adopted to the study of singular transitions a large class of
toric Calabi-Yau fourfolds was constructed in \cite{klyr}. These
fourfolds are constructed as fibrations and thus one might miss
interesting peculiarities of fibration structures when restricting
attention to such a class of objects.

Hence, I decided to base my studies on a class of fourfolds
obtained independently of any fibration structures: the class of
transverse hypersurfaces in weighted projective spaces.

\subsection{Hypersurfaces in weighted projective
spaces}
 \label{weightedhypers} In all dimensions, weighted
projective spaces are toric varieties. Nevertheless, the
relationship between transverse hypersurfaces in weighted
projective spaces and toric Calabi-Yau hypersurfaces in the sense
used in this paper is not a direct one. This is due to the fact,
that the fans of weighted projective spaces are always fans over
the faces of simplices, but in general these simplices are not
reflexive.

The reflexive polytope related to a weighted projective space is
rather given by its maximal Newton polyhedron, i.e. the convex
hull of the homogenous monomials of the appropriate degree to
define a Calabi-Yau hypersurface.

In dimensions less or equal to four, all Newton polyhedra obtained
in this way are reflexive. Though the original weighted projective
spaces are too singular to be members of the corresponding
families of toric varieties\footnote{In particular, generic
hypersurfaces can fail to be $\Sigma$-regular with respect to the
fan of the weighted projective space.}, suitable blow-ups are.

In contrast, in dimensions bigger or equal to five, not all
maximal Newton polyhedra are reflexive, i.e. suitable blow-ups do
not exist. In fact, in five dimensions only about a fifth of all
weight sets allowing transverse polynomials give rise to reflexive
polyhedra.

There are 1.100.055 sets of weights\footnote{Here, the well known
equivalences between sets of weights have of course been modded
out. This is already needed to make the set finite.} allowing for
transverse hypersurfaces. They were calculated in \cite{wisski}
using the algorithm of \cite{klemmschimmrigk} used to classify
transverse threefold hypersurfaces.

Determining the reflexive polyhedra is a conceptually trivial
task. One simply calculates the Newton polyhedra by enumerating
monomials, moves the inner point (all exponents 1) to 0, makes a
change of base to the orthogonal complement of the weight vector
and calculates the convex hull of the resulting point set.
Expressing the convex hull in terms of its bounding halfspaces
makes it a trivial task to check for reflexivity. The only problem
is posed by the huge amount of data. I calculated the convex
hulls\footnote{Since I was also interested in the complete face
structure of the polyhedra, I found it (by experiment) useful not
to use a beneath-beyond type algorithm, but rather a higher
dimensional analogon to the gift-wrapping algorithm. In most cases
where I tried both types of algorithm, the latter (highly
output-sensitive) algorithm proved to be faster.} and in case of
reflexivity also the Hodge numbers by application of the formulae
in \cite{batdualpoly,klyr}. In all cases, they agree with the
numbers found in \cite{wisski} using Landau-Ginsburg methods.

In addition, the reflexive polyhedra were scanned for
equivalences\footnote{One only has to try to find a group element
transforming the polyhedra into each other if all known invariants
are equal. In particular, I used the Hodge numbers and the
individual numbers of faces of different dimensions to perform a
preselection.} modulo $GL(5,\ZZ)$. Though from a physicist's point
of view, Mirror symmetry for fourfolds does not play the same role
as for threefolds, I also checked whether the dual polyhedra also
arise as Newton polyhedra for transverse weights.

The reflexive Newton polyhedra were independently also calculated
by the authors of \cite{kreuzskarkefib4}\footnote{Their results
can be found at
\href{http://hep.itp.tuwien.ac.at/~kreuzer/pub/CY4}{\tt
http://hep.itp.tuwien.ac.at/{$\sim$}kreuzer/pub/CY4}.

Everything found in these lists agrees with my own calculations,
which is a good test for the correctness of both calculations.}.

A short summary of the results is found in table
\ref{transrefltab}. For more details c.f.\ section
\ref{resultsec}.

\begin{table}[ht]
\begin{center}
\begin{tabular}{|lll|rrrr|}
\hline \multicolumn{7}{|c|}{\bf Transverse weights and reflexive
polyhedra (overview)} \\
\hline\hline
\multicolumn{3}{|l|}{transverse weights} &  1.100.055 & 100 \% & & \\
\phantom{-} & \multicolumn{2}{l|}{dual Hodge tripel} & 425.859 & 38,7 \% & & \\
\phantom{-} & \multicolumn{2}{l|}{different Hodge triples} & 667.954 & 60,7 \% & 100 \% & \\
\phantom{-} & \multicolumn{2}{l|}{reflexive polyhedra} & 252.933 & 23,0 \% & & 100 \% \\
\phantom{-} & \phantom{-} & different Hodge triples & 158.178 & & 23,7 \% & 62,5 \% \\
\phantom{-} & \phantom{-} & different mod. $GL(5,\ZZ)$ & 202.746 & & & 80,2 \% \\
\phantom{-} & \phantom{-} & dual Hodge tripel & 90.390 & & & 35,7 \% \\
\phantom{-} & \phantom{-} & dual polyhedron & 31.778 & & & 12,6 \% \\
\hline
\end{tabular}
\parbox{12cm}{\caption[Transverse weights and reflexive polyhedra
(overview)]{{\bf Transverse weights and reflexive polyhedra. }{\em
The entries labelled \emph{``different \ldots''} are lower and
upper bounds for the number of inequivalent families of
hypersurfaces. Equality of Hodge numbers is a necessary and
equivalence of polyhedra a sufficient condition for equality of
the corresponding families. Under \emph{``dual \ldots''} weight
sets are counted, where the dual Hodge triple/polyhedron also
occurs in the same class of weight sets.}}\label{transrefltab}}
\end{center}
\end{table}

\subsubsection{Singularities}
As already mentioned above, the majority of four-dimensional
transverse hypersurfaces in weighted projective spaces are too
singular for their maximal Newton polyhedra to be reflexive. The
simplest example for this is the (well known) family
$$\PP_{1,1,1,1,1,2}[7]$$
of degree 7 hypersurfaces in $\PP_{1,1,1,1,1,2}$.

But even if the Newton polyhedra are reflexive, the fourfolds can
have terminal singularities, i.e. singularities which cannot be
resolved while preserving the canonical class (i.e. without
violating the Calabi-Yau condition). This is related to the fact,
that basic\footnote{i.e. simplices not containing integer points
apart from their vertices} simplices are automatically
elementary\footnote{i.e. having volume 1 in lattice units} in
dimensions 2 and less only. For Calabi-Yau threefolds this is
enough to guarantee smoothness of maximal crepant partial
desingularizations: the relevant cones have dimension 3 and one
additional dimension is ``won'' by reflexivity, which forces the
cones to be index 1 Gorenstein cones\footnote{i.e. $\sigma =
\langle n_1,\ldots,n_k\rangle_{\RR^+}$ and there exists a
primitive integer vector $m$ such that $\forall i:\;\langle
m,n_i\rangle = 1$.}. In higher dimensions, this does not work
anymore.

For a simple example consider the family
$$\PP_{1,1,1,1,2,2}[8].$$
The maximal Newton polyhedron is a reflexive simplex. In
coordinates suitable for the purposes of this section its vertices
are given by
$$\left(\begin{matrix}
 2 & 0 & 0 & 0 & 0 & -1 \\
-1 & 1 & 0 & 0 & 0 & 0 \\
-1 & 0 & 1 & 0 & 0 & 0 \\
 1 & 1 & 1 & 1 & 0 & -2 \\
 0 & 0 & 0 & 0 & 1 & -1
\end{matrix}\right).$$
Now consider the face containing the four vertices corresponding
to the first four columns. Going to the lattice subspace defined
by the face, this face is the simplex with vertices
$$\left(\begin{matrix}
 2 & 0 & 0 & 0 \\
-1 & 1 & 0 & 0 \\
-1 & 0 & 1 & 0
\end{matrix}\right).$$
This is a basic simplex. Therefore, we cannot subdivide the cone
over this face without changing the canonical class and thereby
violating the Calabi-Yau-condition for the generic embedded
hypersurface. On the other hand, the simplex is \emph{not}
elementary --- in lattice units it has volume 2. Thus, the four
vertices cannot be part of a basis for $\ZZ^5$ and we have a
pointlike singularity of the generic hypersurface (the singular
locus in the toric variety is one-dimensional).

\begin{figure}[ht]
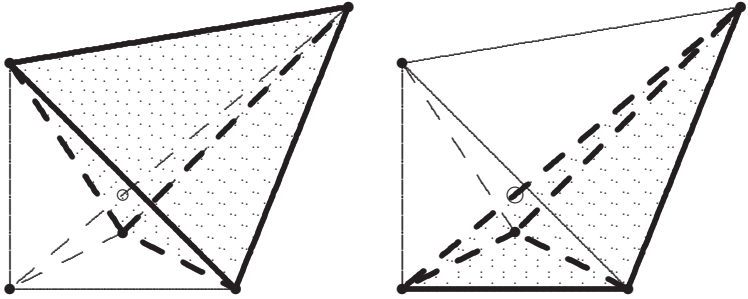

\begin{center}\setlength{\unitlength}{1.5cm}\mbox{
\beginpicture
 \setcoordinatesystem units <\unitlength,\unitlength>
 \setplotarea x from -0.2 to 3.2, y from -0.2 to 2.5
 \put {\circle{0.1}} [Bl] at 1.0 0.83333333
 \put {\circle*{0.1}} [Bl] at 0.0 0.0
 \put {\circle*{0.1}} [Bl] at 2.0 0.0
 \put {\circle*{0.1}} [Bl] at 0.0 2.0
 \put {\circle*{0.1}} [Bl] at 1.0 0.5
 \put {\circle*{0.1}} [Bl] at 3.0 2.5
 \plot 0 2 0 0 2 0 0 2 3 2.5 2 0 /
 \setdashes <0.2\unitlength>
 \plot 1 0.5 0 0 3 2.5 1 0.5 /
 \setsolid
 \setshadegrid span <0.1\unitlength> point at 0 0
 \vshade 0 2 2 <,z,,> 1 0.5 1 <z,,,> 2 0 0 /
 \setshadegrid point at 0.04 0.04
 \vshade 0 2 2 <,z,,> 1 0.5 2.16666667 <z,,,> 3 2.5 2.5 /
 \setshadegrid point at 0.06 -0.04
 \vshade 0 2 2 <,z,,> 2 0 2.33333333 <z,,,> 3 2.5 2.5 /
 \setshadegrid point at 0 0.05
 \vshade 1 0.5 0.5 <,z,,> 2 0 1.5 <z,,,> 3 2.5 2.5 /
 \setplotsymbol ({\circle*{0.05}} [Bl])
 \plot 0 2 2 0 3 2.5 0 2 /
 \setdashes <0.2\unitlength>
 \plot 0 2 1 0.5 2 0 /
 \plot 1 0.5 3 2.5 /
\endpicture}
\mbox{\beginpicture
 \setcoordinatesystem units <\unitlength,\unitlength>
 \setplotarea x from -0.2 to 3.2, y from -0.2 to 2.5
 \put {\circle{0.15}} [Bl] at 1.0 0.83333333
 \put {\circle*{0.1}} [Bl] at 0.0 0.0
 \put {\circle*{0.1}} [Bl] at 2.0 0.0
 \put {\circle*{0.1}} [Bl] at 0.0 2.0
 \put {\circle*{0.1}} [Bl] at 1.0 0.5
 \put {\circle*{0.1}} [Bl] at 3.0 2.5
 \plot 0 2 0 0 2 0 0 2 3 2.5 2 0 /
 \setdashes <0.2\unitlength>
 \plot 0 2 1 0.5 /
 \setsolid
 \setshadegrid span <0.1\unitlength> point at 0 0
 \vshade 0 0 0 <,z,,> 1 0 0.5 <z,,,> 2 0 0 /
 \setshadegrid point at 0.04 0.04
 \vshade 0 0 0 <,z,,> 1 0.5 0.83333333 <z,,,> 3 2.5 2.5 /
 \setshadegrid point at 0.06 -0.04
 \vshade 1 0.5 0.5 <,z,,> 2 0 1.5 <z,,,> 3 2.5 2.5 /
 \setshadegrid point at 0 0.05
 \vshade 0 0 0 <,z,,> 2 0 1.66666667 <z,,,> 3 2.5 2.5 /
 \setplotsymbol ({\circle*{0.05}} [Bl])
 \plot 0 0 2 0 3 2.5 /
 \setdashes <0.2\unitlength>
 \plot 2 0 1 0.5 3 2.5 0 0 1 0.5 /
\endpicture}
\end{center}
\caption{\em The two maximal triangulations of the basic
octahedron. In both cases, one of the simplices is shaded and its
edges are drawn bold. The circular marks in the centers demark the
intersection point of the diagonal with the border of the lower
left simplex. This point is not integer.}\label{octatriangs}
\end{figure}

In the preceding example, the singularity could not be avoided.
The general situation is even more complicated: Different maximal
triangulations (even different projective triangulations) can lead
to different numbers of terminal singularities. In this way,
terminal singularities can occur in phases of models, which also
possess a smooth phase. As a local example for this kind of
phenomenon, consider the cone\footnote{This cone e.g. occurs in
the phase structure of the reflexive hypercube.} $\sigma = \langle
a,b,c,d,e \rangle_{\RR^+}$ with generators
$$a=\left(\begin{matrix}0 \\ 0 \\ 0 \\ 1 \\ 0\end{matrix}\right),\;
b=\left(\begin{matrix}1 \\ 0 \\ 0 \\ 1 \\0\end{matrix}\right),\;
c=\left(\begin{matrix}0 \\ 1 \\ 0 \\ 1 \\0\end{matrix}\right),\;
d=\left(\begin{matrix}0 \\ 0 \\ 1 \\ 1
\\0\end{matrix}\right)\:\:\mbox{and}\:\: e=\left(\begin{matrix}1 \\ 1 \\ 1 \\
1 \\0\end{matrix}\right).$$
 The convex hull of the generators is an octahedron. $\sigma$ has two different
maximal crepant partial desingularizations, which are given by the
two maximal triangulations of this octahedron. The two
triangulations\footnote{Both triangulations are regular, i.e.
allow for a strictly convex piecewise linear function. This, in
turn, guarantees the existence of a K\"ahler form on the resulting
variety.} are depicted in figure \ref{octatriangs}.

The first possibility is to split $\sigma$ into the three cones
$\sigma_1 = \langle a,b,c,e \rangle_{\RR^+}$, $\sigma_2 = \langle
a,c,d,e \rangle_{\RR^+}$ and $\sigma_3 = \langle a,b,d,e
\rangle_{\RR^+}$. As the volume of the octahedron is 3, all three
resulting open patches are smooth.

The second possibility is to split into the two cones $\sigma_1 =
\langle a,b,c,d \rangle_{\RR^+}$ and $\sigma_2 = \langle b,c,d,e
\rangle_{\RR^+}$. While the open patch corresponding to $\sigma_1$
is again smooth, $\sigma_2$ is (up to a change of base) just the
cone from the preceding example and contains a terminal
singularity at its distinguished point.

Although terminal singularities occur for generic members of
families of four-dimensional Calabi-Yau hypersurfaces in toric
varieties, they are of a comparatively mild kind (not only for
hypersurfaces in weighted projective spaces):

\begin{thm}
The terminal singularities of generic members of families of
four-dimensional Calabi-Yau hypersurfaces in toric varieties
corresponding to reflexive polyhedra are at most cyclic quotient
singularities.
\end{thm}

\noindent \textbf{Proof:} The relevant cones for terminal
singularities of generic hypersurfaces are four-dimensional
simplicial cones. Due to reflexivity these are always index 1
Gorenstein cones and can (by choice of base) be written as
$$\sigma = \left\{r \cdot \threevec{x}{1}{0} \mid x\in \Delta \subset
\RR^3, r\in \RR^+\right\},$$
 where $\Delta$ is a basic but not elementary simplex. All of its facets
 are two-dimensional and must hence be elementary in their
 respective lattice subspaces. Therefore one can always choose a
 base such that
 $$\Delta = \mathrm{c.h.}\{\threevec{0}{0}{0},\threevec{1}{0}{0},
 \threevec{0}{1}{0}, \threevec{a}{b}{c}\}$$
with $a,b,c \in \ZZ$, $c > 1$. From the holomorphic quotient
construction one has
$$U(\sigma) = \CC^\star \times \CC^4 / G,$$
where $G = \mathrm{Hom}(A, \CC^\star)$ and $A$ is determined by
the exact sequence
$$0 \longrightarrow \ZZ^4 \stackrel{\alpha}{\longrightarrow} \ZZ^4
\stackrel{\beta}{\longrightarrow} A \longrightarrow 0$$ with
$$\alpha = \left(\begin{array}{cccc}
0 & 0 & 0 & 1 \\ 1 & 0 & 0 & 1 \\ 0 & 1 & 0 & 1 \\ a & b & c & 1
\end{array}\right).$$
Hence, $A = \langle\,\beta(\,(0,0,0,1)\,)\,\rangle = \ZZ_c
\;\Rightarrow\; G = \ZZ_c$.\hfill $\Box$

\subsection{Virtual fibrations} \label{virtfibsearch}
The first step for finding toric Calabi-Yau fibrations is
identifying virtual fibrations. The objects of most interest are
elliptic and K3 fibrations, and thus it would have been desirable
to do a complete scan for both types of virtual fibrations.
Unfortunately, a scan for general K3 fibrations would have
exceeded my possibilities. A search using sublattice enumeration
(c.f. section \ref{sublatalgo}) forbids itself because of its bad
scaling behavior. A search by classification (c.f. section
\ref{classialgo}) is less out of reach, but still exceeded my
possibilities due to the size of the classification list.

For this reason, I concentrated my search on elliptic and elliptic
K3 fibrations. For the elliptic fibrations, the fiber polyhedron
can be one of the 16 reflexive polyhedra\footnote{Mainly for
reference reasons in other lists, a list of these polyhedra may be
found at \cite{frk3fiblist}.} and a search by classification could
be implemented efficiently. The results of this scan can then be
extended to virtual elliptic K3 fibrations using sublattice
enumerations. Due to the known sublattice given by the
two-dimensional subpolyhedron, this extension search now
only\footnote{It obviously also depends on the number of elliptic
subpolyhedra $\Nabla_{ell} \subset \Nabla_{CY}$, which is small in
most cases.} depends linearly on the number of integer points in
${\Nabla_{K3}}$.

As the expense of checking virtual fibrations for obstructions
(c.f. section \ref{obssearch} below) also strongly depends on the
number of points in $\Nabla_{CY}$, I still had to reduce my
program to a subset of all polyhedra identified in section
\ref{weightedhypers}. As a compromise between maximal
exhaustiveness and minimal computational expense for the
calculations on a single polyhedron $\Nabla_{CY}$, I arbitrarily
restricted to polyhedra $\Nabla_{CY}$ with no more than $10,000$
integer points. The restricted set consists of $222,653$
polyhedra, which is roughly 88 \% of the 252,933 polyhedra
corresponding to hypersurfaces in weighted projective spaces.

\subsection{Obstructions for fibrations} \label{obssearch}
Assume we have found a virtual toric elliptic K3 fibration
$$\Nabla_{ell} \subset \Nabla_{K3} \subset \Nabla.$$
How can we find out, whether this gives in fact rise to a
fibration? As noted before, the simple approach of trying all
possible triangulations of $\partial^1\Nabla$ is computationally
much too expensive except for the simplest cases.

Hence, we are looking for a simple way to construct a fan
$\Sigma_b$ and a compatible triangulation of $\partial^1\Nabla$ in
a way that excludes their existence in the case of failure. Where
necessary, we will pose mild additional conditions on the
fibrations, i.e. on $\Sigma_b$ and the triangulation.

\subsection{Constructing projective triangulations}
The general approach in the construction of compatible fans
$\Sigma$ and $\Sigma_b$ will be the following: We first construct
the coarsest fan $\Sigma_b$ meeting a given set of conditions and
then subdivide the fan over codimension $\ge 2$ faces of $\Nabla$
to obtain a fan $\Sigma$ compatible with $\Sigma_b$. At times, we
will further subdivide $\Sigma_b$, which then forces us to further
subdivide $\Sigma$. In the end, any simplicial refinement of
$\Sigma$ will be compatible with $\Sigma_b$.

Since we want to obtain a projective fan $\Sigma$, we must ensure,
that we do not accidentally introduce obstructions against the
existence of stricly convex piecewise linear functions in the
process of subdividing (the normal fan of $\Delta$ allows such
functions).

It is well known, that star-subdividing preserves
projectiveness\footnote{All cones containing the ray $\rho$ around
which one star-subdivides are of the form $\tilde\sigma = \sigma +
\rho$, where $\sigma$ is a cone in the original fan with $\dim
\tilde\sigma = \dim \sigma + 1$. Hence, the linear functions on
these cones given by a strictly convex function on the old fan can
be consistently deformed by changing the value on a generator of
$\rho$ to provide a strictly convex function on the subdivided
fan.}. Unfortunately, this is not what we will be doing. I will
therefore present some additional types of subdivision, which
preserve projectiveness. We will need the following generalization
of a fan:

\begin{defn}
A collection of (not necessarily strictly) convex rational
polyhedral cones is called a \textbf{quasifan}, if the following
holds:
\begin{itemize}
\item[i)] If $\sigma\in\Sigma$ and $\tau \prec \sigma$ is a facet
of $\sigma$, then $\tau \in \Sigma$.
\item[ii)] If $\sigma,\sigma' \in \Sigma$ and $\tau :=
\sigma\cap\sigma'$, then $\tau\prec\sigma$ and $\tau\prec\sigma'$.
\end{itemize}
\end{defn}
The only missing ingredient in comparison with a fan is, that we
do not demand the cones to be strictly convex. Obviously, one can
intersect quasifans to obtain new quasifans just as one can do
with fans. The intersection of a fan with a quasifan is a fan. As
with fans, a quasifan is called projective, if it allows a
strictly convex piecewise linear function.

We need the following almost obvious facts:
\begin{lem} \label{regsub1} Let $\Sigma,\Sigma'$ be two finite
projective quasifans. Then their intersection is a projective
quasifan.
\end{lem}
\textbf{Proof:} The sum of two convex functions is convex. After
scaling one of them (if necessary) the sums of linear pieces
differ whenever the pieces of one summand differ.\hfill$\Box$
\begin{lem} \label{regsub2}
Let $\pi: \QQ^n \twoheadrightarrow \QQ^m$ be a linear map and
$\Sigma$ a regular quasifan with support in $\QQ^m$. Then $\Sigma'
= \{\pi^{-1}\sigma; \sigma\in\Sigma\}$ is a projective quasifan.
\end{lem}
\textbf{Proof:} A piecewise linear strictly convex function $\Psi$
on $\Sigma$ yields the strictly convex function $\Psi \circ \pi$
on $\Sigma'$.\hfill$\Box$

\subsubsection{Codimension 2: K3 fibrations} \label{k3fibcheck}
While finding virtual fibrations is easiest for small dimension of
the fiber, checking for obstructions is easiest for small
codimension of the fiber. We have seen before, that in the case of
codimension 1 of the fiber, no obstructions can occur.
Unfortunately, the minimal fiber codimension in our setting is
two, which is the case I will discuss first.

Here, the fan of the toric base variety lives in the plane, where
a complete fan is uniquely determined by its rays. In addition,
any two-dimensional fan is projective.

I will first discuss triangulations compatible with a flat
prefibration, i.e. a prefibration with pure dimensional
fibers\footnote{For generic defining polynomial, the fibration
fibers will also be pure dimensional.}. One easily sees by
considering restrictions to torus orbits\footnote{This works as
follows: The torus orbits are in 1:1 correspondence to the cones
of the fan, the dimension of an orbit $O_\sigma$ being
$\codim\sigma$. Due to covariance of the torus actions, the
dimension of the fiber over a point in a torus orbit
$O_{\sigma_b}$ of the base is the largest difference between the
dimension of a preimage orbit $O_\sigma$ and the dimension of
$O_{\sigma_b}$. An orbit $O_\sigma, \sigma\in\Sigma$, is mapped
onto the orbit $O_{\sigma_b}$ corresponding to the smallest cone
$\sigma_b$ with $\pi(\sigma) \subset \sigma_b$. For a flat
fibration the dimension of the fiber must not exceed that of the
generic fiber (the other direction is ensured by lemma
\ref{coverbyeqdimface}). Hence, $\dim O_{\sigma_b} + \dim N - \dim
N_b \ge
\dim{O_\sigma}\;\Leftrightarrow\;\dim\sigma\ge\dim{\sigma'}$. This
means, that no cone $\sigma\in\Sigma$ with $\dim\sigma < \dim N_b$
may be mapped into the interior of a higher dimensional cone.
Together with lemma \ref{coverbyeqdimface} this yields the
assertion.}, that this is equivalent to the requirement, that any
cone in $\Sigma$ maps onto a cone in $\Sigma_b$. In particular,
the image of any ray in $\Sigma$ under $\hat\pi$ must be either a
ray of $\Sigma_b$ or the origin. If we want to find a maximal
triangulation of $\partial^1\Nabla$ corresponding to a maximal
crepant desingularization of the total Calabi-Yau variety, our
base fan is now already completely fixed.

It remains to check, whether a complete compatible triangulation
of $\partial^1\Nabla$ exists. Our fan $\Sigma$ (if it exists) will
be a refinement of the fan $\tilde{\Sigma}$ over codimension $\ge
2$ faces of $\Nabla$. In order to be compatible with $\Sigma_b$,
it must be a refinement of the fan $\tilde\Sigma'$ obtained by
intersecting the cones in $\tilde\Sigma$ with the quasifan given
by projection preimages of cones in $\Sigma_b$.  Since $\Sigma_b$
is projective so is $\tilde\Sigma$ due to lemma \ref{regsub1} and
\ref{regsub2}.

For $\Sigma$ to exist, the cones in $\tilde\Sigma'$ must be
generated by integer elements of $\partial^1\Nabla$. If they are,
any complete projective triangulation refining $\tilde\Sigma'$
will be compatible. With lemma \ref{genbyborder} below it suffices
to check intersections with preimages of rays of $\Sigma_b$.

\begin{lem} \label{genbyborder}
Let $A\in\QQ^n$ be an affine hyperplane, $S\subset A$ a finite set
and $\sigma \subset \QQ^n$ an $n$-dimensional strongly convex
polyhedral cone generated by a subset of $S$. Let $\pi: \QQ^n
\rightarrow \QQ^2$ be a linear map and $\sigma_2 \subset \QQ^2$ a
strongly convex cone. Set $\sigma^\cap \defas \sigma \cap \pi^{-1}
\sigma_2$. The following are true:
\begin{itemize}
\item[(i)] Let $\dim \sigma_2 = 2$. Then $\sigma^\cap$ is generated by
elements of $S$ if and only if the intersections $\sigma \cap
\pi^{-1} \tau_1$ and $\sigma \cap \pi^{-1} \tau_2$ are generated
by elements of $S$, where $\tau_1$ and $\tau_2$ are the two facets
(bounding rays) of $\sigma_2$.
\item[(ii)] Let $n > 2$ and $\dim \sigma_2 = 1$. Then $\sigma^\cap$ is
generated by elements of $S$ if and only if $\sigma \cap
\pi^{-1}(0)$ and the intersections $\tau \cap \pi^{-1}\sigma_2$
are generated by elements of $S$ for all proper faces $\tau \prec
\sigma$.
\end{itemize}
\end{lem}
\textbf{Proof:} $\sigma^\cap$ is a strongly convex rational
polyhedral cone. A minimal set of generators must consist of
generators of dimension 1 faces by convexity. In both cases, one
direction follows from the fact, that all faces of $\sigma^\cap$
are generated by subsets of a minimal set of generators for
$\sigma^\cap$. In other words: If $\sigma^\cap$ is generated by a
subset of $S$, so are all of its faces.

For the other directions, first consider case (i): If an element
of a minimal set of generators of $\sigma$ is not contained in
either $\sigma \cap \pi^{-1} \tau_1$ or $\sigma \cap \pi^{-1}
\tau_2$, it must already generate a dimension 1 face of $\sigma$
and hence is a multiple of an element of $S$.

For case (ii), assume a minimal set of generators for
$\sigma^\cap$ contains a vector $\rho$ in the interior of $\sigma$
and $\pi(\rho) \not= 0$. Since $n > 2$, we can find $0 \not=x \in
\QQ^n: \pi(x) = 0$. Consider the line $\rho(t) \defas \rho + t x$,
which is completely contained in $\pi^{-1} \sigma_2$. Since $\rho$
is in the interior of $\sigma$, an open interval around $0$ is
also contained in $\sigma^\cap$. Since $\rho$ and $x$ must be
linearly independent, this contradicts the assumption that $\rho$
generates a dimension 1 face of
$\sigma^\cap$.\hfill$\Box$\vspace{1ex}

Restricting to subspaces containing faces of $\sigma$ and using
complete induction on $n$, we obtain:
\begin{cor}\label{prezeroandtwodim}
Let $A$, $S$, $\sigma$ and $\sigma_2$ be as in lemma
\ref{genbyborder}(ii). $\sigma^\cap$ is generated by elements of
$S$, if and only if $\sigma \cap \pi^{-1}(0)$ and the
intersections $\tau \cap \pi^{-1}\sigma_2$ are generated by
elements of $S$ for all proper faces $\tau \prec \sigma$ with
$\dim \tau \le 2$.
\end{cor}

More general fibrations can only be obtained by omitting rays in
our preliminary $\Sigma_b$. Compatible triangulations can
obviously only be constructed when omitting all rays in $\Sigma_b$
for which the above subcones failed to be generated by integer
elements of $\partial^1\Nabla$. But if the remaining rays still
define a complete fan, i.e. if there are at least two remaining
rays in any halfspace of $N_b$, compatible triangulations are
constructible just as before.

For practical reasons and use in the sequel I will still define
another condition on the prefibration: If the images of the rays
over all vertices of $\Nabla$ are contained in the fan $\Sigma_b$,
the fibration will be called \textbf{singularly flat}. The name is
due to the fact, that such a fibration becomes flat at some
boundary of the family of maximally crepantly desingularized
hypersurfaces corresponding to $\Delta$.

\subsubsection{Codimension 3: Elliptic fibrations}\label{ellfibcheck}
For an elliptic fibration of a Calabi-Yau fourfold the codimension
of the generic fiber and hence the dimension of the base is three.
Thus, the base fan is no longer determined by its rays alone. In
order to use methods similar to the two-dimensional case, we would
have to enumerate the possible sphere triangulations corresponding
to a given set of rays. This is not as bad as enumerating all
possible triangulations of $\partial^1\Nabla$, but in many cases
still very expensive. Even worse, in three dimensions cones are
not forced to be simplicial and for a complete enumeration of
possible base fans we must also consider fans obtained by omitting
lines from a triangulation.

Fortunately, our main interest is not in elliptic fibrations, but
in elliptic K3 fibrations. Here we can use the fact, that the base
fibration
$$\begin{array}{ccc}\PP^1 & & \\
\downarrow & & \\ X_{\Sigma_{b,e}} &
\stackrel{\pi_2}{\longrightarrow} & X_{\Sigma_b,K}\end{array}$$
puts severe restrictions on the possible fans $\Sigma_{b,e}$
compatible with a given fan $\Sigma_{b,K}$: All cones in
$\Sigma_{b,e}$ must project into cones of $\Sigma_{b,K}$.

$\Sigma_{b,e}$ is a complete fan. Writing $N_{b,e} = \ZZ \oplus
N_{b,K}$, it must thus contain the rays generated by $(1,0,0)$ and
$(-1,0,0)$, which constitute the subfan corresponding to the fiber
$\PP^1$. This poses no constraints on the virtual fibrations,
because the preimages of these rays are just the upper and lower
halves of $N_{K3}$ with respect to $N_{ell}$. Now depict the
metric unit sphere in $N_{b,e}$ with north and south pole defined
by the above two rays and subdivided by the meridians defined by
the rays of $\Sigma_{b,K}$ (they are the intersections of the
sphere with the preimages of the rays under $\hat\pi_2: N_{b,e}
\rightarrow N_{b,K}$). Any cone of $\Sigma_{b,e}$ is uniquely
determined by its intersection with the sphere. In order to be
compatible with the K3 fibration, none of the sphere segments
corresponding to cones in $\Sigma_{b,e}$ is allowed to cross the
meridians. The question of regularity of the obtained
triangulations will be postponed to the end of this section.

\subsubsection*{Flat fibrations}
Let us again start by considering flat fibrations. As we are now
dealing with multifibrations, we demand both the elliptic and the
K3 fibration to be flat (which also forces the base fibration to
be flat). The two-dimensional base fan $\Sigma_{b,K}$ for the K3
fibration is uniquely determined as in section \ref{k3fibcheck}.
The fibrations being flat, all rays of $\Sigma_{b,e}$ must pass
through meridians. This is required for the fibration $\pi_2$ to
be flat and already ensured by $\Sigma_{b,K}$ being the fan of the
base of a flat fibration.

Now any compatible flat elliptic fibration $\hat\pi_e: \Sigma
\rightarrow \Sigma_{b,e}$ defines triangulations of the sphere
segments by lemma \ref{coverbyeqdimface} and $\Sigma$ being
simplicial. The vertices of this triangulation are obviously given
by the images of rays over integer points in $\partial^1\Nabla$.
In addition, $\Sigma$ must be given by a refinement of the
partition of $\partial^1\Nabla$ enforced by the K3 fibration, i.e.
it must be a refinement of the fan $\tilde\Sigma'$ from section
\ref{k3fibcheck}. We can now check, whether a flat compatible
elliptic fibration exists: We enumerate triangulations of the
sphere segments and check, whether a compatible refinement of the
initial partition of $\partial^1\Nabla$ exists. The latter can
again be done by calculating intersection cones and testing,
whether they are generated by integer elements of
$\partial^1\Nabla$.

Finding a compatible triangulation of the sphere is significantly
simplified by noting, that the triangulations of the sphere
segments are independent: The preimages of the cones over the
segments are unions of cones in $\tilde\Sigma'$. A triangulation
of one sphere segment induces subdivisons of these cones. The
intersections of cones projecting to neighboring sphere segments
intersect in subcones of the cones projecting to the meridian
between the segments. These subcones are in turn uniquely
determined by the points lying on the meridian. Hence, they are
independent of the two chosen triangulations.

Due to this uniqueness, one should start the search for a
triangulation by checking, whether these subcones are generated by
integer elements of $\partial^1\Nabla$ (if they are not, no
suitable triangulation can exist). For this task, Lemma
\ref{genbyborder} and corollary \ref{prezeroandtwodim} are
applicable after restriction to the planes in $N_{b,e}$ defined by
the meridians and their preimages under $\hat\pi_e$.

The task of enumerating all triangulations of a given sphere
segment can be reduced by noting that the images of cones over
dimension 1 faces of $\Nabla$ must be contained in our
triangulation. This yields an initial partition of the sphere
segments and only refinements thereof have to be enumerated.

\subsubsection*{Singularly flat fibrations}
Let us now consider cases, in which we were unable to construct
base fans and a triangulation compatible with a flat fibration
(which means, that none exists). We want to find out whether a
singularly flat fibration exists. In accordance with the meaning
of the term we demand the elliptic and the K3 fibration to be
singularly flat and the base fibration to be flat\footnote{i.e. we
demand the fibrations to simultaneously become flat at some
boundary of the family of maximally desingularized hypersurfaces}.
Again, our strategy will be to construct the (unique) coarsest
base fan for the elliptic fibration giving rise to a singularly
flat prefibration. First, we consider the K3 fibration. Assume
now, that a fan $\Sigma_{b,K}$ and a compatible triangulation of
$\partial^1\Nabla$ exist. This triangulation is obviously also
compatible with the coarsest possible base fan $\Sigma^c_{b,K}$
leading to a singularly flat fibration, namely the fan defined by
the images under $\hat\pi_K$ of rays over vertices of $\Nabla$.
This is a fan due to reflexivity of $\Nabla$: If a halfspace of
$N_{b,K}$ did not contain the image of a ray over a vertex of
$\Nabla$ the preimage halfspace would contain no vertex. The
latter is a contradiction because $0$ is an interior point of
$\Nabla$.

This coarsest base fan also induces a partition of
$\partial^1\Nabla$ corresponding to the coarsest fan $\Sigma^c$
compatible with $\Sigma^c_{b,K}$. Consider the images of the cones
in $\Sigma^c$ under $\hat\pi_e$. By construction, each of these
images is projected onto exactly one cone of $\Sigma^c_{b,K}$ by
$\hat\pi_2$. The intersection with the sphere is a spheric
polygon, the vertices of which all lie on meridians (either on the
same or on two neighboring meridians). A first necessary condition
for a singularly flat fibration to exist is given by the meridians
themselves, which are subdivided by the points lying on them. The
preimage cones in $\Sigma^c$ must be subdivided to be compatible
with this subdivision and the subcones must be generated by
integer elements of $\partial^1\Nabla$. This can be checked just
as before.

The images of cones over one-dimensional borders of the partition
of $\partial^1\Nabla$ induced by $\Sigma^c$ induce lines on the
sphere between the images of vertices. Lemma
\ref{coverbyeqdimface} ensures, that all necessary two-dimensional
cones in $\Sigma_{b,e}$ are found in this way. If they patch
together to define a (not necessarily simplicial) fan, this is the
coarsest possible fan we sought for, and we only have to check
whether our original partition can be refined to be compatible
with this fan (we do not worry whether finer fans are possible).

How can the lines fail to define a fan? They certainly define a
collection of strongly convex polyhedral cones. The essential
condition is that any two cones intersect along a common face of
both. This can fail in two ways. There could be (i) points on
meridians which fail to emit lines into the neighboring segments
(which would mean a ray in the interior of a 2d cone) or there
could be (ii) lines which intersect inside the sphere segments
(i.e. 2d cones whose intersection is not a face of both). We will
first see that (i) cannot happen and then proceed to treat case
(ii).

Recall how we constructed the fan $\Sigma^c$. The points on
meridians are (scaled) images of vertices of $\Nabla$ or new
vertices of the subdivison of codimension 2 faces of $\Nabla$.
First consider the second case. If a point is not scaled image of
a vertex of $\Nabla$, it must be the scaled image of a point $n$
lying on the intersection of a codimension 2 face with a
hyperplane in $N_\RR$ by lemma \ref{genbyborder} ($n \not\in
\pi_K^{-1}(0)$). Being a vertex of the subface, it must lie in the
interior of an edge of $\Nabla$ by corollary
\ref{prezeroandtwodim}. The two vertices of $\Nabla$ connected by
the edge must lie on different sides of the hyperplane. As the
preimage of the plane in $N_{b,e}$ defined by the meridian (and in
particular of the line through the poles) is contained in the
hyperplane, the cone over the edge passing through the preimage
point must be mapped to the cone over a line crossing the meridian
as desired.

We now turn to points whose scaled preimages are vertices of
$\Nabla$. We will need the following simple consequence of the
duality between faces of $\Nabla$ and $\Delta$:

\begin{lem} \label{edgeinalldir}
Let $v$ be a vertex of $\Nabla$. Any halfspace of $N$ containing
$v$ in its boundary contains an edge $\theta^\star\ni v$ of
$\Nabla$ in its interior.
\end{lem}
\textbf{Proof:} Let $\theta$ be the hypersurface of $\Delta$ dual
to $v$, i.e.\ $\forall m\in\theta: \langle m,v \rangle=-1$. Let
$m^\circ \in \theta^\circ$ be an interior point. Then $M\supset
v^\perp = \RR^+ \{m-m^\circ; m \in \partial \theta\}$. Let the
halfspace be given by $\{n\in N | \langle \hat{m}, n \rangle \le
0\}$ for some $0 \not= \hat{m}\in v^\perp$. Then $\hat{m} = \mu
(m-m^\circ)$ with $\mu \in \RR^{>0}, m \in \partial\theta$. Now $m
\in \partial\theta\;\Rightarrow\; \exists \theta^\star =
\{v+\lambda n; \lambda \ge 0\}$ such that $ \forall
\tilde{n}\in\theta^\star: \langle m, \tilde{n}\rangle = -1$. As
$\lambda>0\;\Rightarrow\;\langle m^\circ, v+\lambda n\rangle >
-1$, one calculates\par\noindent $\qquad\langle \hat{m}, v +
\lambda n \rangle = \mu \langle m-m^\circ, v + \lambda n \rangle =
\mu \left( -1 - \langle m^\circ, v+\lambda n \rangle \right) <
0$.\hfill\vspace{1ex}$\Box$

In particular, consider the two halfspaces separated by the
preimage of the plane on which the meridian lies. Lemma
\ref{edgeinalldir} ensures the existence of two edges starting in
the preimage vertex, which run into the two preimage halfspaces.
By projecting the cones over these edges and intersecting their
images with the sphere we obtain our desired lines.\vspace{1ex}

All lines on the sphere introduced in the above process are
necessary for the resulting fibration to be singularly flat. If
some of them intersect between meridians, we thus have to
introduce a ray passing through the point of intersection. Since
we demand our base fibration to be flat, we also have to introduce
a ray in the base fan for the K3 fibration. In section
\ref{k3fibcheck} we already constructed the finest possible fan
$\Sigma^f_{b,K}$ giving rise to a singularly flat K3 fibration. If
the point of intersection does not project onto a ray of
$\Sigma^f_{b,K}$, we cannot introduce a suitable ray in the base
fan for the K3 fibration, and we have ruled out the existence of a
singularly flat elliptic K3 fibration. Otherwise, we subdivide
$\Sigma_{b,K}$ by introducing the ray and repeat the whole process
for this finer fan. Since there are only finetely many rays we can
add to the K3 fibration's base fan, we only have to repeat the
process finitely many times to either find or exclude the
existence of a singularly flat fibration.

In fact, the subdivison does not introduce new lines on the sphere
except for the new meridian: Consider a fan $\Sigma_{b,K}$ and the
corresponding coarsest compatible fan $\Sigma^c$. We now add a new
ray $\rho_b$ to $\Sigma_{b,K}$ to obtain the refined fan
$\Sigma'_{b,K}$. Let $H^+,H^- \subset N$ be the two halfspaces
separated by $\hat\pi_K^{-1} (\RR \rho_b)$. The coarsest fan
compatible with $\Sigma'_{b,K}$ is then given by
\begin{eqnarray*}
{\Sigma'}^c & = & \{ \sigma \mid \sigma \in
\Sigma^c\;\wedge\;\stackrel{\circ}{\rho_b} \not\subset
\hat\pi_K(\stackrel{\circ}{\sigma}) \} \\ & \cup & \{ \sigma \cap
H \mid H \in \{H^+,H^-\} \wedge \sigma \in
\Sigma^c\;\wedge\;\stackrel{\circ}{\rho_b} \subset
\hat\pi_K(\stackrel{\circ}{\sigma}) \}.
\end{eqnarray*}
Obviously, all two-dimensional cones in ${\Sigma'}^c$ are either
subsets of two-dimensional cones in $\Sigma^c$ or project into
$\rho_b$ under $\hat\pi_K$. Only the latter case induces new lines
on the sphere, which lie on the new meridian. In addition, no
lines on the sphere vanish as a result of the refinement.

Hence, one can simultaneously introduce new rays for all crossing
points and repeat the calculation only once.

\subsubsection*{Even more general fibrations}
As remarked before, checking for all possible base fans is
infeasible for more complex examples. We do not win too much by
using compatibility with the K3 base fan, since we still would
have to consider all K3 fibration base fans given by subsets of
the rays in our above $\Sigma_{b,K}$. Since I want to identify as
many toric fibration structures as possible, I introduce yet
another (weaker) criterion on the fibrations. This allows to
identify a large class of fibrations, which are neither flat nor
singularly flat.

Namely, I require the K3 fibration to be singularly flat and the
base fibration $\pi_2$ to be flat. In other words, the elliptic
fibration does not have to be singularly flat
anymore\footnote{This is somewhat arbitrary and emerged from some
test calculations showing that slightly weaker conditions had much
more effect on the speed of calculation than on the number of
fibrations found.}.

For the base fans this criterion means that all rays of
$\Sigma_{b,e}$ must project to rays in $\Sigma_{b,K}$ under
$\hat\pi_2$. Now consider some fixed $\Sigma_{b,K}$ belonging to a
singularly flat K3 fibration. Note that the sphere segmentation
induced by $\Sigma_{b,K}$ \emph{almost} defines a fan. What is
missing is strict convexity of the cones, i.e. it only defines a
quasifan. The only thing we have to add to obtain a fan is a path
of dividing lines splitting each sphere segment into an upper and
lower half. For a given fan $\Sigma_{b,K}$ it is easy to see (c.f.
the discussion of regularity below), that any fan $\Sigma_{b,e}$
giving rise to a flat base fibration $\pi_2$ must be a refinement
of a fan obtainable in this way.

The points on the meridians through which such a path can possibly
run are given by projections of rays over integer points in
$\partial^1\Nabla$. As before, we denote by $\Sigma^c$ the
coarsest fan compatible with $\Sigma_{b,K}$. The number of points
we have to consider is reduced by first checking, whether all
cones in $\Sigma^c$ projecting to the corresponding rays in
$\Sigma_{b,K}$ can be split accordingly. Then one can check all
lines between points on neighboring meridians for the existence of
a compatible refinement of $\Sigma^c$. Last, one needs to find a
path as described above. This can be done by calculating
connection matrices for the points on neighboring meridians,
multiplying them in circular order and checking for diagonal
elements.

So far, the discussion depends on the choice of $\Sigma_{b,K}$. If
one did not demand the K3 fibration to be singularly flat, one
would have to enumerate all possibilities. The latter basically
consist of enumerating subsets of the set of projections of lines
over integer points in $\partial^1\Nabla$ and can thus be
computationally expensive.

The requirement of a singularly flat K3 fibration reduces the set
of possible fans $\Sigma_{b,K}$ in the following ways: Firstly,
the set of rays $\Sigma^{(1)}_{b,K} \subset \Sigma_{b,K}$ must be
a subset of the rays ${\Sigma^f}^{(1)}_{b,K}$ in the finest
possible fan $\Sigma^f_{b,K}$. Secondly, $\Sigma_{b,K}$ must be a
refinement of the coarsest base fan $\Sigma^c$. This alone reduces
the enumeration to switching on and off elements of
${\Sigma^f}^{(1)}_{b,K} - {\Sigma^c}^{(1)}_{b,K}$.

In fact, one can do even better than this by introducing
cumulative connection matrices for points on the meridians
corresponding to neighboring rays in ${\Sigma^c}^{(1)}_{b,K}$. An
entry of such a matrix is 1 whenever there is a path between the
corresponding points for any subdivision of the corresponding
two-dimensional cone in ${\Sigma^c}^{(2)}$. The existence of a
closed path for any fan $\Sigma_{b,K}$ can then be checked by
calculating all the cumulative connection matrices, multiplying
them and again checking for diagonal elements\footnote{Usually,
one would define the cumulative connection matrices as sums over
products for different subdivisions. The entries would then carry
information about the number of different paths between points.
For the purposes of this paper one only needs to know whether a
path exists or not. The definition given here only keeps this
information, but allows for the additional optimization described
below.}.

Apart from the obvious reduction of the search space, this
approach also has other advantages. The cumulative connection
matrices are calculated as follows: One enumerates all possible
subdivisions of the cone between the corresponding rays. For all
of them, the individual connection matrices are multiplied. The
entries of the cumulative matrix are then 1 whenever the
corresponding entry in any of these products is nonzero. If after
some steps of the enumeration all entries are determined to be 1,
further subdivisions do not have to be considered. If, on the
other hand, all entries are determined to be 0 after enumerating
all subdivisions, further cumulative connection matrices do not
have to be calculated. In this case, no closed path can exist.

Both of these cases occur quite frequently.

\subsubsection*{Regularity of the triangulations}
For flat and singularly flat K3 fibrations we unfortunately cannot
guarantee the enforced partitions to be regular. The situation is
different for the more general fibrations discussed in the last
paragraph. We will see in a moment that they always allow for
compatible regular triangulations.

For flat and singularly flat fibrations we can thereby show that
there is a coarser base fan $\Sigma_{b,e}$, which is compatible
with a regular triangulation of $\partial^1 \Nabla$: Any fan
$\Sigma_b$ as constructed for flat and singularly flat fibrations
is a refinement of a fan as discussed in the last paragraph. All
vertices of the sphere partitions lie on meridians and only
segments of the meridians can end in one of the poles. We start at
some vertex of the sphere partition induced by $\Sigma_{b,e}$,
which is not one of the poles. We then circle around the sphere
following lines which cross the sphere segments. If we can choose
between multiple lines, we always take the upmost one. Since there
are only finitely many lines, we will find a closed path after
finitely many steps.

We now consider simplicial base fans $\Sigma_{b,e}$ with rays
through the poles and one additional ray through all of the
meridians. We will see in a moment, that such a fan is always the
fan over the faces of a convex polytope. The inner face normals of
it then provide the linear pieces of a stricly convex function on
$\Sigma_{b,e}$. As in the two-dimensional case, we use lemma
\ref{regsub2} and \ref{regsub1} to see that at least the coarsest
refinement of the fan over faces with codimension $\ge 2$ of
$\Nabla$ compatible with $\Sigma_{b,e}$ is projective.

In order to see that $\Sigma_{b,e}$ is the fan over a not
necessarily integer\footnote{One can use deformation arguments and
density of $\QQ$ in $\RR$ to find integer polytopes, but for our
purposes this is an unnecessary complication.} convex polytope, we
write $n^+ = (1,0,0)$ and $n^- = (-1,0,0)$ for the poles. We pick
vectors $n^{(i)} \in \rho_i$ for all rays $\rho_i \in \Sigma_b$
such that the projections $\pi_2 n^{(i)}$ lie on the unit circle.
The $\pi_2 n^{(i)}$ obviously form the vertices of a convex
polygon. Now consider the convex hull of all the $n^{(i)}$
together with $r \cdot n^+$ and $r \cdot n^-$, $\RR \ni r \gg 1$.
For sufficiently large $r$, these points are vertices of their
convex hull and $\Sigma_b$ is the fan over its faces.

\subsection{Search results} \label{resultsec}
After the identification of all virtual elliptic K3 fibrations in
the $222,653$ polyhedra $\Nabla_{CY}$ with no more than $10,000$
integer points (c.f.\ section \ref{virtfibsearch}), all of these
fibrations were checked for obstructions as explained above:
First, the virtual K3 fibrations were checked for obstructions
against giving rise to a K3 fibration as described in section
\ref{k3fibcheck}. Depending on the results\footnote{If e.g.\ the
K3 fibration cannot be flat, checking for flat elliptic K3
fibrations makes no sense.}, the full virtual elliptic K3
fibrations were tested as described in section \ref{ellfibcheck}.
In addition, the methods of section \ref{pertgroupcalc} were used
to calculate the perturbative gauge algebras of conjectured
heterotic duals. Lists containing the results can be found at
\cite{frcy4fiblists}.

\subsubsection{Het \texorpdfstring{$\leftrightarrow$}{<->} Het Dualities}
Whenever there are multiple virtual fibration structures within
the fourfold polyhedron, one obtains conjectural dualities between
different dual heterotic string theories. The perturbative gauge
group of one theory is then expected to appear nonperturbatively
in the dual theory.

A particularly important case is given by multiple elliptic K3
fibrations sharing the same elliptic fibration\footnote{The
condition of sharing the same elliptic fibration might be removed
by switching to M-theory compactifications, for which the elliptic
fibration moduli might be unlocked and smooth transitions between
different F-theory limits may be found. Due to the low
dimensionality of the corresponding supersymmetric effective
theories and the corresponding restrictions the predictive power
of such connections is questionable.}. In this setting, one has a
single F-theory compactification defined by the elliptic fibration
and multiple dual heterotic string compactifications via
application of fiberwise duality. Combining such dualities one
gets indirect dualities between the different heterotic string
compactifications.

Among the tested fibration structures, plenty
examples\footnote{There are 92,578 elliptic fibrations
(automorphisms modded out, c.f.\ section \ref{fibcounting})
contained in multiple K3 fibrations.} for multiple fibrations of
this type were found. Due to their special importance they were
collected in separate lists, which may also be found at
\cite{frcy4fiblists}.

\subsubsection{Counting fibrations} \label{fibcounting}
If a reflexive polyhedron $\Nabla_{CY}$ contains multiple virtual
fibrations $\Nabla^{(f)}_1,\Nabla^{(f)}_2 \subset \Nabla_{CY}$,
the natural question arises which of these should be considered as
different. The essential question in this context is, to which
degree virtual fibrations related by automorphisms of
$\Nabla_{CY}$ should be identified.

The most simplistic approach is to just fix a representation of
$\Nabla_{CY}$ and to consider virtual fibrations as different,
whenever the corresponding lattice subspaces differ\footnote{Note
that this already mods out automorphisms of the fiber polyhedra.}.

However, there is no special coordinate system -- neither for the
polyhedra nor the geometric objects they represent. Hence, this
does not appear to be a very natural approach. One should rather
identify virtual fibrations, which are exchanged by an
automorphism of the larger polyhedron $\Nabla_{CY}$, i.e.\ if
$\exists g \in GL(N) \cong GL_n(\ZZ): g \Nabla_{CY} = \Nabla_{CY}
\;\wedge\;g \Nabla^{(f)}_1 = \Nabla^{(f)}_2$.

The lists at \cite{frcy4fiblists} exist in two versions
corresponding to these two ways of counting elliptic fibrations.

The situation is more involved when counting elliptic K3
fibrations because of the special role of elliptic fibrations
shared by multiple elliptic K3 fibrations. Consider a multiple
virtual fibration
$$\begin{array}{ccccc}
& & \Nabla_{K3}^{(1)} & & \\
& \nearrow & & \searrow & \\
\Nabla_{ell} & & & & \Nabla_{CY} \\
& \searrow & & \nearrow & \\
& & \Nabla_{K3}^{(2)} & & \\
\end{array}$$
and an automorphism exchanging the K3 fibrations, i.e. $g \in
GL(N)$ with $g \Nabla_{CY} = \Nabla_{CY}$, $g \Nabla_{ell} =
\Nabla_{ell}$ and $g \Nabla_{K3}^{(1/2)} = \Nabla_{K3}^{(2/1)}$.
The two elliptic K3 fibrations are perfectly equivalent
structures. Each of them gives rise to a conjectured duality
between F-theory and the same heterotic string theory. The
automorphism should thus induce an automorphism of the dual
heterotic string theory exchanging perturbative with
nonperturbative degrees of freedom. For this reason,
identification of the two K3 fibrations is not a desirable
approach and the lists at \cite{frcy4fiblists} contain the
unreduced numbers of K3 fibrations in both versions.

This often leads to a situation in which some K3 polyhedra are not
superpolyhedra of any elliptic polyhedron among a chosen set of
representatives. This peculiarity is not easily resolved. Which of
the K3 polyhedra are superpolyhedra of elliptic polyhedra among a
chosen set of representatives obviously depends on the choice of
representatives. Less obviously but worse, the size of the set of
superpolyhedra also depends on the choice.

Of course, there is always a lower bound on the size of the set of
superpolyhedra, but calculating this bound is computationally
expensive and carries no valuable information.

\setlength{\colsepsik}{\tabcolsep}\setlength{\tabcolsep}{0.7
\tabcolsep}
\begin{table}[ht]
\begin{center}
\begin{tabular}{|ll|rl|rl|}
\hline\multicolumn{6}{|c|}{\textbf{Numbers of fibration structures}} \\
\hline\hline \multicolumn{2}{|l|}{reflexive polyhedra} & 252,933 &
\multicolumn{3}{l|}{} \\\hline
 \multicolumn{2}{|l|}{with $\le 10,000$ integer points} & 222,653 & \multicolumn{3}{l|}{(88.0 \%)}
\\\hline
 \multicolumn{2}{|l|}{with virtual elliptic fibration} &
215,877 & \multicolumn{3}{l|}{(97.0 \%)} \\\hline
 \phantom{-} &
{with virt. ell. K3 fibration} &
 200,157 & \multicolumn{3}{l|}{(89.9 \% / 92.7 \%)} \\\hline\hline
\multicolumn{2}{|l|}{virtual elliptic fibrations} & 634,827 & &
488,788 &
\\\hline \multicolumn{2}{|l|}{virtual K3 fibrations} &
842,661 & \multicolumn{3}{l|}{} \\\hline
 & flat K3 fibrations & 40,404 & \multicolumn{3}{l|}{(4.8 \%)}  \\\hline
 & sing. flat K3 fibrations & 266,092 & \multicolumn{3}{l|}{(31.6 \%)} \\\hline
\multicolumn{2}{|l|}{virtual elliptic K3 fibrations} & 1,783,067 &
& 1,126,791 & \\\hline
 & flat elliptic fibration & 19,424 & (1.1 \%) & 16,381 & (1.5 \%) \\\hline
 & sing. flat ell. fibration & 67,531 & (3.8 \%) & 53,890 & (4.8 \%) \\\hline
 & more general & 523,776 & (29.4 \%) & 395,950 & (35.1 \%)
\\\hline
\end{tabular}
\parbox{12cm}{\caption[Numbers of fibration structures]{{\textbf{Numbers of fibration structures.}}
{\em For elliptic and elliptic K3 fibrations the two listed
numbers refer to different ways of counting (c.f.\ section
\ref{fibcounting}). The numbers for K3 fibrations refer to virtual
elliptic K3 fibrations neglecting the elliptic fibration
structure. Fibrations fulfilling stronger criteria are not
contained in the numbers for weaker criteria, i.e. a flat
fibration is not counted as singularly flat.}}\label{fibovv} }
\end{center}
\end{table}
\setlength{\tabcolsep}{\colsepsik}

\section{Monodromy - two series of examples}
In section \ref{monodromy1} we saw, that in many cases the gauge
algebra of a dual heterotic theory cannot be deduced by
considering the K3 polyhedron alone.

Among the 1,126,791 virtual elliptic K3 fibrations\footnote{The
numbers refer to fibrations after identification of elliptic
fibrations by automorphisms of $\Nabla$.} found, 619,059 K3
polyhedra give rise to nontoric divisors, which might potentially
lead to a non-simply-laced gauge group by monodromy. In slightly
more than half of these cases (310,396) there actually is a
nontrivial action of monodromy on the exceptional fibers leading
to non-simply laced gauge algebras. In almost as many cases
(279,856) no identifications occur. The discrepancy between the
numbers is given by cases, where the (nontoric) section of the
elliptic fibration of the K3 fibers patches together to a
multisection rather than a section of the total elliptic
fibration\footnote{This happens in 28,807 of the 129,437 examples
with nontoric section.}.

Assuming that fourfold polyhedra do not systematically prefer or
suppress the occurrence of identifications, this was to be
expected\footnote{The possibility of independent identifications
in the upper and lower half of a polyhedron in some cases
increases the probability to find \emph{some} identification.}.

That either possibility should be suppressed was not to be
expected, because one can easily construct series of examples for
both possibilities. The latter are much simpler than the examples
found in my search and yet provide a huge set of examples, if one
wishes to concentrate on the question of monodromy. Therefore, I
will briefly discuss their construction.

\subsection{Monodromy}
Examples for cases with complete identification by monodromy are
given by the simplest examples one can imagine: trivial
fibrations. Of course, only the K3 fibration can be trivial.
Otherwise we would obtain no non-abelian gauge group at all. Let
$\Delta_K$ be a reflexive three-dimensional polyhedron together
with a complete triangulation of $\partial \Nabla_K$, and
$\Delta_2$ a two-dimensional reflexive polyhedron. The direct
product of the corresponding toric varieties is given by the
product fan which in turn corresponds to the reflexive polyhedron
$\Delta = \Delta_K \times \Delta_2$. The dual $\Nabla$ is the
convex hull of $(\Nabla_K, 0)$ and $(0, \Nabla_2)$ which obviously
has $\Nabla_K$ as a reflexive subpolyhedron, and we do not need to
worry about the virtual toric prefibration being a fibration. It
is also clear, that all faces of $\Nabla_K$ are also faces of
$\Nabla$ with the same dimension. Hence, no face of $\Nabla_K$ can
be in the interior of a codimension 2 face of $\Nabla$, and
identifications via monodromy occur whenever possible. As a side
effect, though, nontoric sections of the K3 fibers' elliptic
fibrations patch together to multisections rather than sections of
the total elliptic fibration.

\subsection{No Monodromy}
Examples for this case can easily be constructed using the other
(dual) extreme method for constructing reflexive polyhedra. By
choosing $\Nabla = \Nabla_K \times \Nabla_2$ all faces of
$\Nabla_K$ lie on faces of $\Nabla$ with equal \emph{codimension}.
A straightforward but lengthy calculation shows that all of these
virtual toric prefibrations define toric fibrations. In addition,
one can always find fans such that the K3 fibration becomes flat.
Note though that one can only find fans for flat elliptic
fibrations in special cases\footnote{Flat fibrations can be
obtained if and only if the height (in integer units) of all
points in $\Nabla_K$ above or below the elliptic plane is less
than or equal to $1$.}.

\appendix
\section{Algorithms for calculating lattice bases}
\label{completetobase} In this section, I will give constructive
proofs for two lemmata used for the construction of bases for
sublattices of $\ZZ^d$.

\begin{lem} \label{onecomplete} Let $B_1\in\ZZ^d$ be a primitive vector. Then there
exist matrices $B, I \in GL(d,\ZZ)$ with
$$B = \left(\begin{array}{ccc}
\rule{0.1pt}{0.5cm} & & \rule{0.1pt}{0.5cm} \\
B_1 & \cdots & B_d \\
\rule{0.1pt}{0.5cm} & & \rule{0.1pt}{0.5cm}
\end{array}\right) \qquad\mbox{and}\qquad B^{-1} = I =
\left(\begin{array}{rcl}
\rule[0.5ex]{0.5cm}{0.1pt} & I_1 & \rule[0.5ex]{0.5cm}{0.1pt} \\
& \vdots & \\
\rule[0.5ex]{0.5cm}{0.1pt} & I_d & \rule[0.5ex]{0.5cm}{0.1pt}
\end{array}\right).$$
\end{lem}
\textbf{Proof:} Using the Euclidean algorithm, we can construct a
map $C:\ZZ^d\rightarrow\ZZ^d$ such that $\forall x\in\ZZ^d:
\langle x,C(x) \rangle = \mathrm{g.c.d.}(x)$. First set $I_1
\defas C(B_1)$. For $n=2,\ldots,d$ we then construct $I_n, B_n$
such that $$\forall i\le n: \langle B_i, I_n\rangle = \langle B_n,
I_i\rangle = \delta_{i,n}.$$

To this end choose a primitive $I_n \in
\{B_1,\ldots,B_{n-1}\}^\perp$. Set $\tilde{B}_n \defas C(I_n)$ and
finally $$B_n \defas \tilde{B}_n - \sum_{i=1}^{n-1} \langle
\tilde{B}_n, I_i \rangle B_i.$$ The assertion follows after
finitely many steps.\hfill$\Box$

\begin{lem} \label{multicomplete} Let $X\defas (x_1 \cdots x_n) \in \mathrm{Mat}(\ZZ; d, n)$
be an integer matrix. Then $\exists \overline{B}\defas (b_1 \cdots
b_d) \in GL(d,\ZZ)$ such that
$$\mathrm{rk}\, (x_1\cdots x_r) = s\quad \Rightarrow \quad \langle x_1,\ldots, x_r
\rangle_\QQ = \langle b_1,\ldots, b_s \rangle_\QQ.$$
\end{lem}
\textbf{Proof:} Without loss of generality we can assume the $x_i$
to span $\QQ^d$ (otherwise complete with e.g. some vector base of
the orthogonal complement). We further assume $\forall i: x_i
\not=0$. We use complete induction on the dimension $d$. The case
$d = 1$ is trivially solved by $(1)$. Assume that the assertion is
true for $d-1$. Then set $B_1 \defas x_1 / \mathrm{g.c.d.}(x_1)$
and calculate matrices $B, I$ as in Lemma \ref{onecomplete}.
Define the matrix $X' \in \mathrm{Mat}(\ZZ; d-1, n)$ by
$$I\,X \asdef
\left(\begin{array}{rcl}
\rule[0.5ex]{0.5cm}{0.1pt} & \star & \rule[0.5ex]{0.5cm}{0.1pt} \\
& X' & \end{array}\right).$$ Using the assumption we calculate
$\overline{B}' \in GL(d-1,\ZZ)$ and set
$$\overline{B} \defas B\;
\left(\begin{array}{cc@{}c@{}ccc@{}c}
1 & & & 0 & \cdots & 0 & \\
\cline{3-6}
0 & & \vline & & & & \\
\vdots & & \vline & & \overline{B}' & & \\
0 & & \vline & & & &
\end{array}\right).$$
\hfill$\Box$

\section{K3 Picard lattices} \label{K3latticeformulae}
The following is a short summary of the formulae used to calculate
the Picard lattice of a generic toric K3 hypersurface. A
derivation and proofs may be found in \cite{latticek3}.

Let $\Delta \subset M$ be a reflexive three-dimensional
polyhedron, $N \supset \Nabla = \Delta^\ast$. Denote by $Z$ the
generic smooth element of the corresponding family of K3 surfaces.
The Picard number of $Z$ is then given by $$\rho(Z) = l(\Delta) -
4 - \sum_{\codim\, \theta^\ast = 1} l^\ast(\theta^\ast) +
\sum_{\codim\, \theta^\ast = 2}
l^\ast(\theta)l^\ast(\theta^\ast),$$ where the $\theta$
($\theta^\ast$) are faces of $\Delta$ ($\Nabla$), $l(X)$ denotes
the number of integer points contained in $X$ and $l^\ast(X)$ the
number of integer points in the relative interior. Note that
\emph{generic} is a stronger condition than in similar statements
for higher dimension: The Picard number is strictly larger for a
dense subset of the space of defining polynomials.

For $n \in N \cap \partial^1\Nabla$ denote by $D_n$ the
intersection of the corresponding T-Weil divisor with $Z$, which
will be called a toric divisor. $D_n$ is irreducible except for
the following situation:
\begin{itemize}
\item[(i)] $n$ lies in the interior of an edge $\theta^\ast$ of
$\Nabla$.
\item[(ii)] The length of the dual edge $\theta$ of $\Delta$
(measured in integer units) is larger than 1.
\end{itemize}
In this case, $D_n$ splits into $l(\theta) = |m_1 - m_2|$
irreducible components
$$D_n = \sum_{i=1}^{l(\theta)} \tilde{D}_n^{(i)}.$$
 Here $l(\theta)$ denotes the integer length of $\theta$, and
$m_1, m_2$ are the two vertices of $\Delta$ at the boundary of
$\theta$. The irreducible components of $D_n$ are called semitoric
\footnote{in contrast to the divisors not equivalent to divisors
with support in $X_\Delta - (\CC^\star)^3$, which emerge at a
dense subset of the parameter space} divisors.

The nonvanishing intersections between divisors corresponding to
different $n,n'$ are:
$$\begin{array}{rcll}
D_n \cdot D_{n'} & = & l(\theta) & \mbox{if $n, n'$ are neighbors on a common edge $\theta^\star$} \\
\tilde{D}_n^{(i)} \cdot D_{n'} & = & 1 & \mbox{if $n'$ is a vertex and $n$ is a neighbor in the interior of} \\
& & & \mbox{a common edge} \\
\tilde{D}_n^{(i)} \cdot \tilde{D}_{n'}^{(j)} & = & \delta_{i,j} &
\mbox{if $n, n'$ are neighbors in the interior of a common edge}
\end{array}$$

The self-intersections are determined by the general rule
$$D_n \cdot D_n = \sum_{n'} \langle m,n' \rangle D_n \cdot D_{n'},$$
where $m\in M$ such that $\langle m,n \rangle = -1$ (e.g. a normal
of some face on which $n$ lies) and the sum ranges over all
neighbors $n'$ of $n$ on common edges.

In particular, for $n$ in the interior of some edge of $\Nabla$,
one obtains
$$\begin{array}{rcl}
D_n \cdot D_n & = & -2 l(\theta) \\
\tilde{D}_n^{(i)} \cdot \tilde{D}_n^{(i)} & = & -2. \end{array}$$

\section*{Acknowledgements}
I would like to thank W.\ Nahm, K.\ Wendland and D.\ Roggenkamp
for helpful discussions and comments. Computing facilities for the
calculations described in this paper were mainly provided by \\
\phantom{x} $\bullet$ Physikalisches Institut der Universit\"at Bonn \\
\phantom{x} $\bullet$ MPI f\"ur Mathematik, Bonn \\
\phantom{x} $\bullet$ Konrad-Adenauer-Stiftung, Sankt Augustin \\
I would also like to thank several people, who provided computer
hardware but do not want to be named.



\begin{thebibliography}{BKMT99}
\setlength{\parskip}{0.6ex}

\bibitem[AKM00]{liecy3andf}
\textsc{P.~S. Aspinwall, S.~Katz und D.~R. Morrison},
\textsl{{Lie} groups,
  {Calabi-Yau} threefolds, and {F}-theory},
\newblock Adv. Theor. Math. Phys. \textbf{4}, 95--126 (2000),
  \href{http://xxx.lanl.gov/abs/hep-th/0002012}{hep-th/0002012}.

\bibitem[AKMS97]{searchk3fib}
\textsc{A.~Avram, M.~Kreuzer, M.~Mandelberg und H.~Skarke},
\textsl{Searching
  for {K3} Fibrations},
\newblock Nucl. Phys. \textbf{B494}, 567--589 (1997),
  \href{http://xxx.lanl.gov/abs/hep-th/9610154}{hep-th/9610154}.

\bibitem[AL96]{needk3fordual}
\textsc{P.~S. Aspinwall und J.~Louis}, \textsl{On the Ubiquity of
{K3}
  Fibrations in String Duality},
\newblock Phys. Lett. \textbf{B369}, 233--242 (1996),
  \href{http://xxx.lanl.gov/abs/hep-th/9510234}{hep-th/9510234}.

\bibitem[Aud91]{audin}
\textsc{M.~Audin},
\newblock The Topology of Torus Actions on Symplectic Manifolds,
\newblock in \textsl{Progress in Math.}, volume~93, Birkh\"auser, 1991.

\bibitem[Bat93]{batyrevholquot}
\textsc{V.~V. Batyrev},
\newblock Quantum cohomology rings of toric manifolds,
\newblock in \textsl{Journ\'ees de G\'eom\'etrie Alg\'ebrique d'Orsay (Juillet
  1992)}, volume Ast\'erisque 218, pages 9--34, Soci\'et\'e Math\'ematique de
  France, 1993.

\bibitem[Bat94]{batdualpoly}
\textsc{V.~V. Batyrev}, \textsl{Dual Polyhedra and Mirror Symmetry
for
  {Calabi-Yau} Hypersurfaces in Toric Varieties},
\newblock J. Alg. Geom. \textbf{3}, 493--535 (1994),
  \href{http://xxx.lanl.gov/abs/alg-geom/9310003}{alg-geom/9310003}.

\bibitem[BB97]{batbordualcones}
\textsc{V.~V. Batyrev und L.~A. Borisov},
\newblock Dual cones and mirror symmetry for generalized {C}alabi-{Y}au
  manifolds,
\newblock in \textsl{Mirror Symmetry, II}, edited by B.~Greene und S.-T. Yau,
  pages 71--86, Providence, RI, 1997, Amer. Math. Soc.,
  \href{http://xxx.lanl.gov/abs/alg-geom/9402002}{alg-geom/9402002}.

\bibitem[BCOG00]{cy4andf}
\textsc{V.~Braun, P.~Candelas, X.~D.~L. Ossa und A.~Grassi},
\textsl{Toric
  {Calabi-Yau} fourfolds, duality between {N = 1} theories and divisors that
  contribute to the superpotential},
\newblock (2000),
  \href{http://xxx.lanl.gov/abs/hep-th/0001208}{hep-th/0001208}.

\bibitem[BKMT99]{candowithb2}
\textsc{P.~Berglund, A.~Klemm, P.~Mayr und S.~Theisen}, \textsl{On
Type {IIB}
  Vacua With Varying Coupling Constant},
\newblock Nucl. Phys. \textbf{B558}, 178--204 (1999),
  \href{http://xxx.lanl.gov/abs/hep-th/9805189}{hep-th/9805189}.

\bibitem[Bor86]{k3diffeo2}
\textsc{C.~Borcea}, \textsl{Diffeomorphisms of a {K3} surface},
\newblock Math. Ann. \textbf{275}, 1--4 (1986).

\bibitem[BPS99]{candowithb1}
\textsc{M.~Bershadsky, T.~Pantev und V.~Sadov}, \textsl{{F}-Theory
with
  Quantized Fluxes},
\newblock Adv. Theor. Math. Phys. \textbf{3}, 727--773 (1999),
  \href{http://xxx.lanl.gov/abs/hep-th/9805056}{hep-th/9805056}.

\bibitem[BSW97]{bsw02mirr}
\textsc{R.~Blumenhagen, R.~Schimmrigk und A.~Wi{\ss}kirchen},
\textsl{(0,2)
  mirror symmetrie},
\newblock Nucl. Phys. \textbf{B486}, 598 (1997),
  \href{http://xxx.lanl.gov/abs/hep-th/9609167}{hep-th/9609167}.

\bibitem[CF98]{groupvisibleinpoly}
\textsc{P.~Candelas und A.~Font}, \textsl{Duality between the webs
of heterotic
  and {Type II} vacua},
\newblock Nucl. Phys. \textbf{B511}, 295--325 (1998),
  \href{http://xxx.lanl.gov/abs/hep-th/9603170}{hep-th/9603170}.

\bibitem[Cox95]{coxholquot}
\textsc{D.~A. Cox}, \textsl{The homogenous coordinate ring of a
toric variety},
\newblock J. Algebraic Geom. \textbf{4}, 17--50 (1995).

\bibitem[Cox97]{coxrecent}
\textsc{D.~A. Cox},
\newblock Recent Developments in Toric Geometry,
\newblock in \textsl{Algebraic geometry --- Santa Cruz 1995}, edited by
  J.~Kollar et~al., pages 389--436, Providence, RI, 1997, Amer. Math. Soc.,
  \href{http://xxx.lanl.gov/abs/alg-geom/9606016}{alg-geom/9606016}.

\bibitem[Don90]{k3diffeo3}
\textsc{S.~Donaldson}, \textsl{Polynomial Invariants For Smooth
  Four-Manifolds},
\newblock Topology \textbf{29}, 257--315 (1990).

\bibitem[Ful93]{fulton}
\textsc{W.~Fulton},
\newblock \textsl{Introduction to Toric Geometry},
\newblock Springer-Verlag, Berlin Heidelberg New York, 1993.

\bibitem[GSO76]{gso1}
\textsc{F.~Gliozzi, J.~Scherk und D.~Olive}, \textsl{Supergravity
and the
  spinor dual model},
\newblock Phys. Lett. \textbf{65B} (1976).

\bibitem[GSO77]{gso2}
\textsc{F.~Gliozzi, J.~Scherk und D.~Olive},
\textsl{Supersymmetry,
  supergravity theories and the dual spinor model},
\newblock Nucl. Phys. \textbf{B122} (1977).

\bibitem[KLRY98]{klyr}
\textsc{A.~Klemm, B.~Lian, S.-S. Roan und S.-T. Yau},
\textsl{{Calabi-Yau}
  fourfolds for {M}- and {F}-Theory compactifications},
\newblock Nucl. Phys. \textbf{B518}, 515--574 (1998),
  \href{http://xxx.lanl.gov/abs/hep-th/9701023}{hep-th/9701023}.

\bibitem[KS94]{klemmschimmrigk}
\textsc{A.~Klemm und R.~Schimmrigk}, \textsl{{Landau-Ginzburg}
string vacua},
\newblock Nucl. Phys. \textbf{B411}, 559--583 (1994),
  \href{http://xxx.lanl.gov/abs/hep-th/9204060}{hep-th/9204060}.

\bibitem[KS98a]{kreuzskarkefib4}
\textsc{M.~Kreuzer und H.~Skarke}, \textsl{Calabi-Yau 4-folds and
toric
  fibrations},
\newblock J. Geom. Phys. \textbf{26}, 272--290 (1998),
  \href{http://xxx.lanl.gov/abs/hep-th/9701175}{hep-th/9701175}.

\bibitem[KS98b]{kreuzskarkek3list}
\textsc{M.~Kreuzer und H.~Skarke}, \textsl{Classification of
Reflexive
  Polyhedra in Three Dimensions},
\newblock Adv. Theor. Math. Phys. \textbf{2}, 847--864 (1998),
  \href{http://xxx.lanl.gov/abs/hep-th/9805190}{hep-th/9805190}.

\bibitem[KS00]{kreuzscarkeclass4poly}
\textsc{M.~Kreuzer und H.~Skarke}, \textsl{Complete classification
of reflexive
  polyhedra in four dimensions},
\newblock (2000),
  \href{http://xxx.lanl.gov/abs/hep-th/0002240}{hep-th/0002240}.

\bibitem[LSW99]{wisski}
\textsc{M.~Lynker, R.~Schimmrigk und A.~Wi{\ss}kirchen},
  \textsl{{Landau-Ginzburg}-Vacua of string-, {M-} and {F}-theory at c=12},
\newblock Nucl. Phys. \textbf{B550}, 123--150 (1999),
  \href{http://xxx.lanl.gov/abs/hep-th/9812195}{hep-th/9812195}.

\bibitem[Mat85]{k3diffeo1}
\textsc{T.~Matumoto},
\newblock On Diffeomorphisms of a {K3} surface,
\newblock in \textsl{Algebraic and Topological theories --- to the memory of
  Dr. Takehiko Miyaka}, edited by M.~Nagata et~al., pages 616--621, Kinokuniya,
  Tokyo, 1985.

\bibitem[Mus94]{musson}
\textsc{I.~Musson}, \textsl{Differential operators on toric
varieties},
\newblock J. Pure and Appl. Alg. \textbf{95}, 303--315 (1994).

\bibitem[Nik80]{nikulinlemma}
\textsc{V.~V. Nikulin}, \textsl{Finite automorphism groups of
{Kaehler} {K3}
  surfaces},
\newblock Trans. Mosc. Math. Soc. \textbf{38}, 71--135 (1980).

\bibitem[PS97]{k3chowformula}
\textsc{E.~Perevalov und H.~Skarke}, \textsl{Enhanced Gauge
Symmetry in Type
  {II} and {F}-Theory Compactifications: {Dynkin} Diagrams From Polyhedra},
\newblock Nucl. Phys. \textbf{B505}, 679--700 (1997),
  \href{http://xxx.lanl.gov/abs/hep-th/9704129}{hep-th/9704129}.

\bibitem[Roha]{frcy4fiblists}
\textsc{F.~Rohsiepe},
\newblock List of hypersurfaces in weighted projective spaces, reflexive Newton
  polyhedra and fibration structures,
\newblock \\\href{http://www.th.physik.uni-bonn.de/People/rohsiepe
  }{http://www.th.physik.uni-bonn.de/People/rohsiepe}.

\bibitem[Rohb]{frk3fiblist}
\textsc{F.~Rohsiepe},
\newblock Lists of elliptic toric K3 surfaces, sections and gauge algebras,
\newblock
  \href{http://www.th.physik.uni-bonn.de/People/rohsiepe}{http://www.th.physik%
.uni-bonn.de/People/rohsiepe}.

\bibitem[Roh04]{latticek3}
\textsc{F.~Rohsiepe}, \textsl{Lattice polarized toric {K3}
surfaces},
\newblock (2004),
  \href{http://xxx.lanl.gov/abs/hep-th/0409290}{hep-th/0409290}.

\end{thebibliography}
\end{document}